\documentclass[11pt,preprint]{aastex}
\pdfoutput=1
\usepackage{graphicx}
\usepackage{lscape}
\newcommand\msun {M_\odot}
\newcommand\mearth {{M_\oplus}}
\newcommand\gtsima{$\; \buildrel >\over\sim \;$}
\newcommand\simgt{\lower.5ex\hbox{\gtsima}}
\newcommand\ltsima{$\; \buildrel <\over\sim \;$}
\newcommand\simlt{\lower.5ex\hbox{\ltsima}}
\newcommand\piEbold{{\mathbold \pi_E}}

\newcommand\rep{{\tilde r}_E}
\newcommand{\mathbold}[1]{\mbox{\boldmath $\bf#1$}}

\shorttitle{}
\shortauthors{Bennett et al}

\begin{document}

\title{
Discovery and Mass Measurements of a Cold, 10-Earth Mass Planet
and Its Host Star
}

\author{Y.~Muraki$^{1,67}$, 
C.~Han$^{2,68,\ast}$, 
D.P.~Bennett$^{3,67,70,\ast}$,
D.~Suzuki$^{4,67}$,
L.A.G.~Monard$^{5,68}$, \\
R.~Street$^{6,69}$,
U.G.~Jorgensen$^{7,71}$, 
P.~Kundurthy$^{8}$,
J.~Skowron$^{9,68}$,
A.C.~Becker$^{8}$, \\
M.D.~Albrow$^{10,70}$, 
P.~Fouqu\'e$^{11,70}$, 
D.~Heyrovsk\'y$^{12}$,
R.K.~Barry$^{13}$,
J.-P.~Beaulieu$^{14,70}$,  \\
D.D.~Wellnitz$^{15}$,
I.A.~Bond$^{16,67}$,
T.~Sumi$^{4,17,67}$, 
S.~Dong$^{18,68}$,
B.S.~Gaudi$^{9,68}$, \\
D.M.~Bramich$^{19,69}$,
M.~Dominik$^{20,21,69,71}$,\\
and \\
F.~Abe$^{4}$, 
C.S.~Botzler$^{22}$, 
M.~Freeman$^{22}$, 
A.~Fukui$^{4}$, 
K.~Furusawa$^{4}$, 
F.~Hayashi$^{4}$, \\
J.B.~Hearnshaw$^{10}$,
S.~Hosaka$^{4}$, 
Y.~Itow$^{4}$,  
K.~Kamiya$^{4}$, 
A.V.~Korpela$^{23}$,
P.M.~Kilmartin$^{24}$, \\
W.~Lin$^{16}$,
C.H.~Ling$^{16}$, 
S.~Makita$^{4}$, 
K.~Masuda$^{4}$,  
Y.~Matsubara$^{4}$, 
N.~Miyake$^{4}$, 
K.~Nishimoto$^{4}$, \\
K.~Ohnishi$^{25}$, 
Y.C.~Perrott$^{22}$,
N.J.~Rattenbury$^{26}$, 
To.~Saito$^{27}$,
% T.~Sako$^{4}$, 
L.~Skuljan$^{16}$, \\ 
D.J.~Sullivan$^{23}$, 
W.L.~Sweatman$^{16}$,
P.J.~Tristram$^{24}$,
K. Wada$^{1}$, 
P.C.M.~Yock$^{22}$, \\ (The MOA Collaboration)\\
G.W.~Christie$^{28}$,
D.L.~DePoy$^{29}$, 
E.~Gorbikov$^{30}$,
A.~Gould$^{9}$,
S.~Kaspi$^{30}$, 
C.-U.~Lee$^{31}$,  \\
F.~Mallia$^{32}$,
D.~Maoz$^{30}$,
J.~McCormick$^{33}$,
D.~Moorhouse$^{34}$,
T.~Natusch$^{28}$,
B.-G.~Park$^{31}$,\\
R.W.~Pogge$^{9}$, 
D.~Polishook$^{35}$,
A.~Shporer$^{30}$,
G.~Thornley$^{34}$,
J.C.~Yee$^{9}$,\\
(The $\mu$FUN Collaboration)\\
A.~Allan$^{36}$, 
P.~Browne$^{20,71}$,
K.~Horne$^{20}$,
N.~Kains$^{19}$, \\ 
C.~Snodgrass$^{37,38,71}$,
I.~Steele$^{39}$, 
Y.~Tsapras$^{6,40}$, \\
(The RoboNet Collaboration)\\ 
V.~Batista$^{14}$,
C.S.~Bennett$^{41}$, 
S.~Brillant$^{37}$, 
J.A.R.~Caldwell$^{42}$,
A.~Cassan$^{14}$,
A.~Cole$^{43}$, \\
R.~Corrales$^{14}$,
Ch.~Coutures$^{14}$,
S.~Dieters$^{43}$, 
D.~Dominis Prester$^{44}$,
J.~Donatowicz$^{45}$, \\
J.~Greenhill$^{43}$, 
D.~Kubas$^{14,37}$,
J.-B.~Marquette$^{14}$,
R.~Martin$^{46}$,
J~Menzies$^{47}$, \\
K.C.~Sahu$^{48}$,
I.~Waldman$^{49}$,
A.~Williams$^{46}$  
M.~Zub$^{50}$, \\ (The PLANET Collaboration)\\
H.~Bourhrous$^{51}$,
Y.~Matsuoka$^{52}$,
T.~Nagayama$^{52}$,
N.~Oi$^{53}$,
Z.~Randriamanakoto$^{47}$,\\ (IRSF Observers) \\
V.~Bozza$^{54,55}$,
M.J.~Burgdorf$^{56,57}$,
S.~Calchi Novati$^{54}$, 
S.~Dreizler$^{58}$,
F.~Finet$^{59}$, 
M.~Glitrup$^{6}$,\\
K.~Harps{\o}e$^{7}$,
% F.V.~Hessman$^{58}$,
T.C.~Hinse$^{7,31}$,
M.~Hundertmark$^{58}$,
C.~Liebig$^{20}$, 
G.~Maier$^{50}$, \\
L.~Mancini$^{54,61}$, 
M.~Mathiasen$^{7}$,
S.~Rahvar$^{62}$,
D.~Ricci$^{59}$,
G.~Scarpetta$^{54,55}$, 
J.~Skottfelt$^{7}$,\\
J.~Surdej$^{59}$, 
J.~Southworth$^{63}$,
J.~Wambsganss$^{50}$, 
F.~Zimmer$^{50}$,\\ (The MiNDSTEp Consortium)\\
A.~Udalski$^{64}$, 
R.~Poleski$^{64}$, 
{\L}.~Wyrzykowski$^{64,65}$, 
K.~Ulaczyk$^{64}$, 
M.K.~Szyma{\'n}ski$^{64}$,  \\
M.~Kubiak$^{64}$,
G.~Pietrzy{\'n}ski$^{64,66}$, 
I.~Soszy{\'n}ski$^{64}$ \\ (The OGLE Collaboration)\\
}


%\title{Supplementary Information}
%\author{}
%\date{}
%\maketitle

\altaffiltext{1}{Department of Physics, Konan University, Nishiokamoto 8-9-1, Kobe 658-8501, Japan}
\altaffiltext{$\ast$}{To whom correspondence should be addressed; E-mail: bennett@nd.edu,
cheongho@chungbuk.ac.kr}
\altaffiltext{2}{Department of Physics, Chungbuk National University, 410 Seongbong-Rho, Hungduk-Gu, Chongju 371-763, Korea}
\altaffiltext{3}{Department of Physics, 225 Nieuwland Science Hall, University of Notre Dame, Notre Dame, IN 46556, USA}
\altaffiltext{4}{Solar-Terrestrial Environment Laboratory, Nagoya University, Nagoya, 464-8601, Japan}
\altaffiltext{5}{Bronberg Observatory, Centre for Backyard Astrophysics, Pretoria, South Africa}
\altaffiltext{6}{Las Cumbres Observatory Global Telescope Network, 6740 Cortona Dr.,
Suite 102, Goleta, CA 93117, USA}
\altaffiltext{7}{Niels Bohr Institute and Centre for Stars and Planet Formation, Juliane Mariesvej 30, 2100 Copenhagen, Denmark} 
\altaffiltext{8}{Astronomy Department, University of Washington, Seattle, WA 98195} 
\altaffiltext{9}{Department of Astronomy, Ohio State University, 140 West 18th Avenue, Columbus, OH 43210, USA}
\altaffiltext{10}{University of Canterbury, Department of Physics and Astronomy, Private Bag 4800, Christchurch 8020, New Zealand}
\altaffiltext{11}{IRAP, CNRS, Universit\'{e} de Toulouse, 14 avenue Edouard Belin, 31400 Toulouse, France}
\altaffiltext{12}{Institute of Theoretical Physics, Charles University, V 
Hole\v{s}ovi\v{c}k\'ach 2, 18000 Prague, Czech Republic}
\altaffiltext{13}{Goddard Space Flight Center, Greenbelt, MD 20771, USA}
\altaffiltext{14}{Institut d'Astrophysique de Paris, F-75014, Paris, France}
\altaffiltext{15}{University of Maryland, College Park, MD 20742, USA}
\altaffiltext{16}{Institute for Information and Mathematical Sciences, Massey University, Private Bag 102-904, Auckland 1330, New Zealand}
\altaffiltext{17}{Department of Earth and Space Science, Osaka University, 1-1 Machikaneyama-cho, Toyonaka, Osaka 560-0043, Japan}
\altaffiltext{18}{Sagan Fellow; Institute for Advanced Study, Einstein Drive, Princeton, NJ 08540, USA}
\altaffiltext{19}{European Southern Observatory, Karl-Schwarzschild-Stra{\ss}e 2, 85748 Garching bei M\"{u}nchen, Germany}
\altaffiltext{20}{SUPA, University of St Andrews, School of Physics \& Astronomy,North Haugh, St Andrews, KY16 9SS, UK}
\altaffiltext{21}{Royal Society University Research Fellow}
\altaffiltext{22}{Department of Physics, University of Auckland, Private Bag 92-019, Auckland 1001, New Zealand}
\altaffiltext{23}{School of Chemical and Physical Sciences, Victoria University, Wellington, New Zealand}
\altaffiltext{24}{Mt. John University Observatory, P.O. Box 56, Lake Tekapo 8770, New Zealand}
\altaffiltext{25}{Nagano National College of Technology, Nagano 381-8550, Japan}
\altaffiltext{26}{Jodrell Bank Observatory, The University of Manchester, Macclesfield, Cheshire SK11 9DL, UK}
\altaffiltext{27}{Tokyo Metropolitan College of Aeronautics, Tokyo 116-8523, Japan}
\altaffiltext{28}{Auckland Observatory, P.O. Box 24-180, Auckland, New Zealand}
\altaffiltext{29}{Department of Physics, Texas A\&M University, 4242 TAMU, College Station, TX 77843-4242, USA}
\altaffiltext{30}{School of Physics and Astronomy, Raymond and Beverley Sackler Faculty of Exact Sciences, Tel-Aviv University, Tel Aviv 69978, Israel}
\altaffiltext{31}{Korea Astronomy and Space Science Institute, 776 Daedukdae-ro, Yuseong-gu 305-348 Daejeon, Korea}
\altaffiltext{32}{Campo Catino Austral Observatory, San Pedro de Atacama, Chile}
\altaffiltext{33}{Farm Cove Observatory, 2/24 Rapallo Place, Pakuranga, Auckland 1706, New Zealand}
\altaffiltext{34}{Kumeu Observatory, Kumeu, New Zealand}
\altaffiltext{35}{Benoziyo Center for Astrophysics, Weizmann Institute of Science}
\altaffiltext{36}{School of Physics, University of Exeter, Stocker Road, Exeter, EX4 4QL, UK}
\altaffiltext{37}{European Southern Observatory, Casilla 19001, Vitacura 19, Santiago, Chile}
\altaffiltext{38}{Max-Planck-Institut fŸr Sonnensystemforschung, Katlenburg-Lindau, Germany}
\altaffiltext{39}{Astrophysics Research Institute, Liverpool John Moores University, Twelve Quays House, Egerton Wharf, Birkenhead CH41 1LD, UK}
\altaffiltext{40}{Astronomy Unit, School of
Mathematical Sciences, Queen Mary, University of London, London E1 4NS}
\altaffiltext{41}{Department of Physics, Massachusetts Institute of Technology, 77 Mass. Ave., Cambridge, MA 02139}
\altaffiltext{42}{McDonald Observatory, 16120 St Hwy Spur 78 \#2, Fort Davis, TX 79734}
\altaffiltext{43}{University of Tasmania, School of Mathematics and Physics, Private Bag 37, Hobart, TAS 7001, Australia}
\altaffiltext{44}{Department of Physics, University of Rijeka, Omladinska 14, 51000 Rijeka, Croatia}
\altaffiltext{45}{Technische Universitaet Wien, Wieder Hauptst. 8-10, A-1040 Wienna, Austria} 
\altaffiltext{46}{Perth Observatory, Walnut Road, Bickley, Perth 6076, WA, Australia} 
\altaffiltext{47}{South African Astronomical Observatory, P.O. Box 9 Observatory 7925, South Africa}
\altaffiltext{48}{Space Telescope Science Institute, 3700 San Martin Drive, Baltimore, MD 21218, USA}
\altaffiltext{49}{University College London, Dept. of Physics and Astronomy, Gower Street, London WC1E 6BT, UK}
\altaffiltext{50}{Astronomisches Rechen-Institut, Zentrum f\"ur Astronomie der Universit\"at Heidelberg, M\"onchhofstrasse 12-14, 69120 Heidelberg, Germany}
\altaffiltext{51}{Department of  
Mathematics and Applied Mathematics, University of Cape Town,  
Rondebosch 7701,  Cape Town, South Africa}
\altaffiltext{52}{Graduate School of Science, Nagoya University, Furo- 
cho, Chikusa-ku, Nagoya 464-8602, Japan}
\altaffiltext{53}{Department of Astronomical Science, The Graduate  
University for Advanced Studies (Sokendai), Mitaka, Tokyo 181-8588,  
Japan}
\altaffiltext{54}{Department of Physics, University of Salerno, Via Ponte Don Melillo, 84084 Fisciano (SA), Italy}
\altaffiltext{55}{Istituto Nazionale di Fisica Nucleare, Sezione di Napoli, Italy}
\altaffiltext{56}{SOFIA Science Center, NASA Ames Research Center, Mail Stop N211-3, Moffett Field CA 94035, USA }
\altaffiltext{57}{ Deutsches SOFIA Institut, Universitaet Stuttgart, Pfaffenwaldring 31, 70569 Stuttgart, Germany}
\altaffiltext{58}{ Institut fur Astrophysik, Georg-August-Universitat, Friedrich-Hund-Platz 1, 37077 Gottingen, Germany}
\altaffiltext{59}{ Institut dÕAstrophysique et de Geophysique, Allee du 6 Aout 17, Sart Tilman, Bat. B5c, 4000 Liege, Belgium}
\altaffiltext{60}{ Department of Physics \& Astronomy, Aarhus University, Ny Munkegade 120, 8000 Arhus C, Denmark}
\altaffiltext{61}{ Max Planck Institute for Astronomy, K\"onigstuhl 17, 69117 Heidelberg, Germany}
\altaffiltext{62}{ Department of Physics, Sharif University of Technology, and School of Astronomy, IPM, 19395-5531, Tehran, Iran}
\altaffiltext{63}{ Astrophysics Group, Keele University, Staffordshire, ST5 5BG, United Kingdom}
\altaffiltext{64}{Warsaw University Observatory, Al.~Ujazdowskie~4, 00-478~Warszawa, Poland}
\altaffiltext{65}{Institute of Astronomy, Univ.~of Cambridge, Madingley Road, Cambridge CB3 0HA, UK}
\altaffiltext{66}{Universidad de Concepci{\'o}n, Departamento de Astronomia, Casilla 160--C, Concepci{\'o}n, Chile}
\altaffiltext{67}{Microlensing Observations in Astrophysics (MOA) Collaboration}
\altaffiltext{68}{Microlensing Follow Up Network ($\mu$FUN) Collaboration}
\altaffiltext{69}{RoboNet Collaboration} 
\altaffiltext{70}{Probing Lensing Anomalies Network (PLANET) Collaboration} 
\altaffiltext{71}{Microlensing Network for the Detection of Small Terrestrial Exoplanets (MiNDSTEp) Collaboration}

\clearpage

\begin{abstract}
We present the discovery and mass measurement 
of the cold, low-mass planet MOA-2009-BLG-266Lb, made with the
gravitational microlensing method. This planet has a mass
of $m_p = 10.4 \pm 1.7{\it\mearth}$ and orbits a star
of mass $M_\star = 0.56\pm 0.09{\it\msun}$ at a 
semi-major axis of $a = 3.2{+1.9\atop -0.5}\,$AU and an orbital period
of $P = 7.6{+7.7\atop -1.5}\,$yrs. The planet and host star mass measurements are
enabled by the measurement of the microlensing parallax effect, which is 
seen primarily in the light curve distortion 
due to the orbital motion of the Earth. But, the 
analysis also demonstrates the capability to measure microlensing parallax
with the Deep Impact (or EPOXI) spacecraft in a Heliocentric orbit.
The planet mass and orbital distance are similar to predictions for
the critical core mass needed to accrete a substantial gaseous envelope,
and thus may indicate that this planet is a ``failed" gas giant. 
This and
future microlensing detections will test planet formation theory predictions
regarding the prevalence and masses of such planets.
\end{abstract}

\section{Introduction}
\label{sec-intro}

In the leading core accretion planet formation model \citep{lissauer_araa}, a key
role is played by the ``snow line", where the proto-planetary disk
becomes cold enough for ices to condense. The timescale for 
agglomeration of small bodies into protoplanets is shortest just
beyond the snow line, because this is where the
surface density of solid material is highest. The largest
protoplanets in these regions are expected to quickly reach a mass of $\sim
10M_\oplus$ by accumulating the majority of the
solid material in their vicinity.  They then slowly 
accrete a gaseous envelope of hydrogen and helium. The envelope
can no longer maintain hydrostatic equilibrium when it reaches the mass
of the core, so it collapses, starting a rapid gas accretion phase that
leads to a massive giant planet.  The hydrostatic accretion phase is
predicted to have a much longer duration than the other two phases of
solid accretion and rapid gas accretion \citep{pollack96}.  This has 
several possible implications, including a higher
frequency of low-mass, rocky/icy planets than gas
giants, a feature in the final mass function of planets near the
critical core mass of $\sim 10~M_\oplus$, a relative paucity of
planets with masses of $10-100~M_\oplus$ \citep{idalin04}, and the
formation of very few gas giants orbiting low-mass hosts  \citep{laughlin04},
where the gas disks are expected to dissipate before the critical core
mass is reached.

These predictions follow from general physical considerations, but
they also rely upon a number of simplifying assumptions
that make the calculations tractable. So, they could be incorrect. For
example, recent work suggests that uncertainties in the
initial surface density of solids in the protoplanetary disk, 
grain opacities in protoplanetary atmospheres, and 
the size distribution of accreting planetesimals 
can radically alter the timescales of these various phases and 
thus the resulting distribution
of final planet masses \citep{rafikov11,hubickyj05,movshovitz08}.  
Therefore, the measurement of the mass function of
planets down to below the predicted critical core mass will provide
important constraints on the physics of planet formation.

Attempts to test core accretion theory predictions with the mass
distribution of the $\sim 500$ detected exoplanets and the
$\sim 1200$ candidate exoplanets found by the Kepler mission
\citep{borucki11} have met with varied success.
Radial velocity detections confirm the prediction that massive
gas giants should be rare around low-mass stars \citep{johnson10}, 
but the  prediction that $10-100~M_\oplus$ planets 
should be rare in short period orbits is contradicted by the data \citep{howard10}.
Kepler finds a large population of planets at
$\sim 2.5\,R_\oplus$ in short period orbits, which is
consistent with a result from the radial velocity planet detection 
method \citep{howard10}. This might be considered a confirmation of the
core accretion theory prediction that $\sim 10\,\mearth$ ``failed gas giant
core" planets should be common, but in fact all of the low-mass planets
found by radial velocity and transit methods have been well interior to
the snow line, where these ``failed core" planets are thought to form.
It is possible that the exoplanet mass (or radius) function is quite
different outside the snow line due to such processes as
sorting by mass through migration \citep{ward97} and photo-evaporation 
of gaseous envelopes \citep{baraffe05}. Thus, a study of the exoplanet
mass function beyond the snow line should provide a sharper test
of the core accretion theory.

The gravitational microlensing method 
\citep{mao91,bennett_rev,gaudi_rev} has demonstrated sensitivity extending down
to planets of mass $< 10\mearth$ in orbits beyond the snow line
\citep{bennett96,ogle390,bennett08}.
Thus it can provide a complementary probe
of the physics of planet formation for planets that have migrated
little from their putative birth sites.
A statistical analysis of some of the first
microlensing discoveries \citep{gould10} indicates that cold, Saturn-mass
planets beyond the snow line are more common than gas giants found
in closer orbits with the Doppler radial velocity method \citep{cumming08}.
Another microlensing study \citep{sumi10} of the mass function slope
showed that planets of
$\sim 10\mearth$ are even more common than these cold Saturns,
in general agreement with the core accretion theory prediction for 
``failed gas giant cores" \citep{kennedy_snowline,thommes08}.

A well sampled planetary microlensing light curve provides a direct
determination of the planet:star mass ratio, but not the individual
masses of planet and host star. 
Furthermore, planets found by microlensing typically have distant, low-mass
host stars, so their
faintness makes them difficult to characterize.
While finite source effects in the light curve do constrain a combination
of the mass and distance, it has often been necessary to estimate the planet
and host masses and distance with a Bayesian analysis 
\citep{ogle390} based on a Galactic model and prior assumptions
about the exoplanet distribution. When the masses of planetary
microlenses and their host stars can be measured, they
will provide tighter constraints on planet formation theory.

\begin{figure}
\centering
\includegraphics[width=15cm]{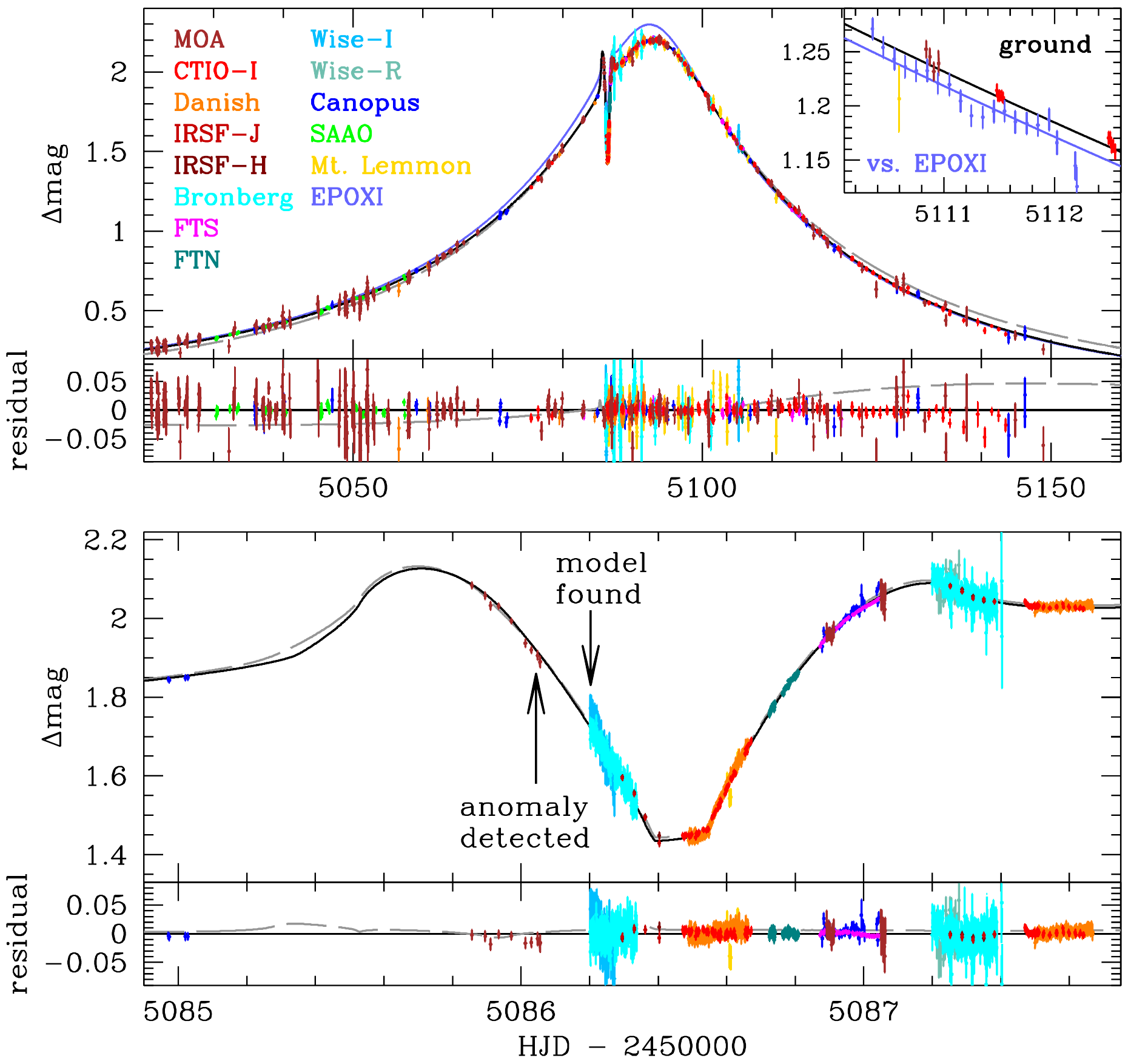}
\caption{\noindent Data and best-fit model of the MOA-2009-BLG-266
microlensing event plotted with respect to magnitude of the unmagnified source. 
The upper panel shows the part of the microlensing light curve magnified by more
than 25\% and the lower panel shows a close-up of the planetary deviation. The
sub-panel at the bottom of each panel shows the residual to the best fit model,
which is given by the solid black curves. The light-blue curve in the top panel
is the model light curve for the position of the EPOXI spacecraft and the inset
shows the data binned EPOXI photometry from the $\sim 2\,$day period of
observations from the EPOXI/HRI instrument. The grey-dashed curve in
each panel is the best fit non-parallax microlensing model.
}
\label{fig-lc}
\end{figure}

Here, we present the first example of a mass measurement for a
cold, low-mass planet discovered by microlensing, which has a
mass very similar to the expected critical mass for gas accretion.
The light curve of microlensing event MOA-2009-BLG-266 exhibits a
planetary signal due to a companion with a mass ratio of
$\sim 6 \times 10^{-5}$ (see Figure~\ref{fig-lc}). The 
light curve also reveals a microlensing parallax signal due to
the orbital motion of the 
Earth \citep{gould-par1,macho-par1}. 
When  combined with the information from the finite size of the source
during the planetary perturbation, this allows one to work out the complete
geometry of the microlensing event \citep{gould-par1}, yielding a
measurement of the host and planet masses. This has been done previously at
this level of precision only for the giant planets in the system
OGLE-2006-BLG-109L \citep{gaudi-ogle109,bennett-ogle109}. 
In addition, observations from the EPOXI
spacecraft in a heliocentric orbit
% \citep{refsdal-par,dong-ogle05smc1} 
corroborate the mass measurement for MOA-2009-BLG-266Lb, and 
demonstrate the potential of obtaining masses for planetary events
that are too brief for a parallax measurement due to the Earth's orbit.

Our observations are described in Section~\ref{sec-obs}, and
Section~\ref{sec-data} details our data reduction procedures. 
Section~\ref{sec-src_prop} presents a 
detailed discussion of the source star properties, and we 
discuss the determination of the properties of the planetary system in
Section~\ref{sec-pl_char_mod}. 
Finally, we discuss some of the implications of this discovery
in Section~\ref{sec-conclude}.

\section{Observations}
\label{sec-obs}

The microlensing event MOA-2009-BLG-266 
[$({\rm RA},{\rm DEC})=(17^{\rm h}\ 48^{\rm m}\ 05.95^{\rm s}, 
-35^\circ \ 00'\ 19.48'')$ and $(l,b)=(-4.9^\circ,-3.6^\circ)$]
was discovered on 1 June 2009 by the Microlensing Observations
in Astrophysics (MOA) collaboration
MOA-II 1.8m survey telescope at Mt.\ John University Observatory 
in New Zealand.  The Probing Lensing Anomalies NETwork (PLANET),
Microlensing Follow-Up Network ($\mu$FUN) and Microlensing Network for
the Detection of Small Terrestrial Exoplanets (MiNDSTEp)
teams followed some of the early part of the light curve.
The wide field of view ($2.2\,{\rm deg}^2$) of the MOA-II survey telescope 
allows MOA to monitor the Galactic bulge with a high enough cadence
to discover planetary signals in any of the 500-600 microlensing events
they discover every year, and this has resulted in the discovery
of several exoplanets \citep{sumi10,bennett08}. On 11 September 2009,
the MOA survey detected such an anomaly in the MOA-2009-BLG-266
light curve and announced it as a probable planetary anomaly.
In response to the alert, the event was intensively observed using the
combined telescopes of the $\mu$FUN, PLANET, RoboNet, and
MINDSTEp teams, resulting in nearly complete light curve coverage
for the last $\sim 75$\% of the anomaly. 
Within four hours, modeling by MOA confirmed that
this was almost certainly a planetary event, which led to
observations by the IRSF infrared telescope 
at the South African Astronomical Observatory (SAAO). This and
further modeling by MOA and $\mu$FUN, as well as rapid reduction
of $\mu$FUN data
prevented observing resources from being diverted to other interesting
events that were found the same day.

Our data set consists of observations from 13 different telescopes, with
several telescopes contributing data in different passbands. We treat each
telescope-passband combination as an independent data set with independent
flux parameters in the microlensing light curve fits, and this combined data
set includes 18 telescope-passband combinations. The planetary signal was first
seen in data from the MOA-II 1.8m survey telescope \citep{sako_moacam3}
at Mt.\ John University Observatory in New Zealand. This analysis includes
1996 MOA-II observations taken from 2007-2009.

In response to the MOA-II microlensing event alert on 1 June 2009 and the
microlensing anomaly alert on 11 September 2009, data were obtained from a
number of follow-up groups. The PLANET
collaboration \citep{ogle390}
added this event to its target list for the 1.0m telescope at Mt.\ Canopus 
Observatory near Hobart, Australia, and the 1.0m telescope at the
South African Astronomical Observatory (SAAO) on 16 July 2009 as a regular
planet search target following the alert plus follow-up planet detection
strategy \citep{gouldloeb92}. Unfortunately, the PLANET observing time allocation
at SAAO
ended on 18 August 2009, which was nearly a month prior to the planetary signal.
Additional observations prior to the detection of the planetary
anomaly were also obtained from $\mu$FUN-CTIO, MiNDSTEp-Danish,
and Robonet-Faulkes South. These observations help to constrain the
microlensing parallax signal, but the parallax signal is primarily detected
in the MOA data.

Canopus had 35 observations spanning 49 days prior to the planetary
signal, including four observations on the night prior to the beginning
of the planetary signal. MOA had no observations on the two days prior
to the beginning of the planetary anomaly, so the Canopus data were the
only observations taken on the night before the planetary anomaly
began. These observations indicated no deviation from a single
lens light curve, and this indicated that the anomaly had a very
short duration, as is typical for light curve deviations due to low-mass
planets. Thus, the Canopus data contributed to the identification
of the planetary nature of the anomaly identified in the MOA data.
This was 
important because another anomalous event and a high magnification
event were also identified by MOA on the same day.
The final data set contains 205 $I$ band observations from Canopus and
33 $I$ band observations from SAAO.

The first data in response to MOA's 11 September 2009 alert on the
planetary anomaly came from the Microlensing Follow-up Network
($\mu$FUN) with observations from the 0.4 m telescope Bronberg Observatory 
in Pretoria, South Africa and the 1.0 m telescope of Wise Observatory in Israel,
which were able to begin observations $\sim 4\,$hours after the 
anomaly alert (which coincided with the last MOA observation on the 
night of the alert). About two hours later (after the MOA planetary
light curve model had been circulated), a series of observations
were begun with the 1.4m InfraRed Survey Telescope (IRSF), 
which is located at SAAO and
features simultaneous imaging in the $J$, $H$,and $K$ bands.
The final data set includes 597 unfiltered observations from Bronberg,
36 and 30 observations in $I$ and $R$ (respectively) from Wise,
and 19, 20, and 18 observations in the $J$, $H$, and $K$ bands 
from IRSF. The raw Bronberg data consist primarily of very densely sampled
observations during the two nights of the planetary deviation, and the 1705
observations from Bronberg were binned to a 7.2 minute cadence to yield the
597 measurements that were used for all the light curve modeling. The IRSF
observations are much sparser, but they do include coverage of the
first caustic crossing endpoint, as well as observations from March 2010,
when the microlensing magnification was $< 1.01$.

The $\mu$FUN group also obtained a large number of observations
in the $H$, $I$, and $V$ bands from the ANDICAM instrument
on the 1.3 m SMARTS telescope at CTIO in Chile. This instrument
observes simultaneously in the optical and infrared, so the final
data set includes 861 $H$ band observations, which mostly 
overlap in time with the 317 $I$ band and 56 $V$ band observations
that are included in the final data set. The CTIO data also include
regular sampling of the stellar microlensing light curve
after the planetary anomaly and a few images from 2010, so they
contribute significantly to the microlensing parallax constraints. 
The Microlensing Network for the Detection of Small Terrestrial Exoplanets
(MiNDSTEp) also obtained dense light curve coverage of the planetary 
deviation using the 1.54 m Danish Telescope at the European Southern Observatory
in La Silla, Chile, and their data set consists of 611 $I$ band observations.
Another $\mu$FUN telescope in the Americas was the
1.0 m telescope at Mt.\ Lemmon Observatory in Arizona
which contributed 73 $I$ band observations to the final data set.

The rise of the light curve from the planetary magnification ``trough"
was covered largely by the 2.0m Faulkes telescopes operated by the
Las Cumbres Global Telescope Network (LCOGTN). The Faulkes
North telescope (FTN) located in Haleakala, Hawaii contributed 
148 SDSS-$I$ band observations to the final data set, while the
Faulkes South Telescope (FTS) located in Siding Springs,
Australia, contributed 128 SDSS-$I$ band 
observations to the final data set. The Canopus and MOA telescopes
also covered the last part of the rise from the light curve ``trough\rlap."

The Robonet group also obtained a substantial amount of FTS data in the
SDSS-$g$, $r$ and Pan Starrs-$z$ and $y$ with 52, 51, 49, and 115
images in each passband. This multi-color data was obtained because
it was thought that it might be helpful to help calibrate the unfiltered
EPOXI data.

Our complete light curve
data set also includes 929 OGLE-III $I$-band observations that
ended on 3 May 2009 with the termination of the OGLE-III survey, when
the magnification was $A\approx 1.06$. (OGLE-III was terminated to enable
an upgrade to a more sensitive camera with a larger field of view for the
OGLE-IV survey.)

Finally, we obtained high angular resolution
AO images from the NACO instrument on the 
European Southern Observatory's Very Large Telescope (VLT) facility
in 2010 after the event was over.

% Real-time modeling of the MOA survey discovery data plus pre-discovery
% Canopus data indicated the planetary origin of the perturbation, with the
% first planetary models circulated just at the time of the first follow-up observations
% from Wise and Bronberg. In addition, the relatively 
% long time scale of the event raised the need for extended follow-up 
% observations to measure the microlens parallax, which enables complete 
% determination of the physical parameters of the lens when combined with 
% the Einstein radius that is measurable from the perturbation.  As a 
% result, observations were conducted until the second week of November 
% when the event was difficult to observe due to its proximity to the sun.

\section{Data Reduction}
\label{sec-data}

Most of the photometry was done using the 
Difference Image Analysis (DIA) method \citep{tom96}. The MOA images
were reduced with the MOA DIA pipeline \citep{bond01}, while the PLANET,
RoboNet-II, MINDSTEp and most 
of the $\mu$FUN data were reduced with a DIA routine following the
same basic strategy as ISIS \citep{ala98}, but using a numerical kernel \citep{Bramich2008}.
The implementation of this numerical kernel DIA routine that was used for
most of the data was pySIS (v3.0) \citep{albrow09}
but the Robonet pipeline was used for the FTS data \citep{Bramich2008}. The OGLE data were
reduced with the OGLE pipeline \citep{ogle-pipeline}.
The Mt.\ Lemmon data were reduced
with DoPHOT \citep{dophot}, and the IRSF data were reduced with SoDoPHOT,
which was derived from DoPHOT \citep{bennett-sod}. SoDoPHOT was also
used to reduce the CTIO $I$ and $V$ band data, but this SoDoPHOT reduction
was only used to help calibrate the EPOXI photometry. The multicolor FTS data and
the CTIO data were also reduced with ALLFRAME \citep{allframe} to aid the
EPOXI photometry calibration, but only the SoDoPHOT reductions of the
CTIO $I$ and $V$ band data were used in the final EPOXI calibrations.
The pySIS reductions of the
CTIO data were used for light curve modeling.

One difficulty that is sometimes encountered with DIA photometry is that 
excess photometric scatter can result for images where the target star is much
brighter or much fainter than it is in the reference frame. This effect was noticed
in the pySIS reductions of the Canopus data. So, the final Canopus photometry
was a combination of two reductions based on reference frames in which
the brightness of the target was very different. The relative normalization of these
two reductions was determined by a linear fit with the 3-$\sigma$ outliers removed 
from each data set. Then, the final Canopus photometry was determined by a weighted
sum of these two data sets with the weighting determined by the difference between
the target brightness in the image being reduced and the two reference frames.

An additional correction is necessary for the
unfiltered Bronberg data. The color dependence of atmospheric extinction
can give rise to systematic photometry errors because the color of the
source star is typically slightly different from the color of the stars used to
normalize the photometry. This gives rise to a photometry error that scales
as the airmass for monochromatic light and a static atmosphere. For a very
wide passband, like that of Bronberg, the effective passband depends on the
amount of atmospheric extinction, so the photometric error does not follow
the scaling with airmass very precisely  \citep{stubbs07}. Furthermore, the
amount of dust in the atmosphere can change with time. So, we correct the
Bronberg photometry by normalizing the photometry of the target star to a set
of stars with a similar color \citep{bennett-ogle109}.

The VLT/NACO data were reduced following the procedures used for the
analysis of MOA-2007-BLG-192 \citep{moa192_naco}.

\subsection{Reduction of EPOXI Data}
\label{sec-epoxi}

For a period of just under three days on 10-12 October 2009, we were
able to obtain observations using the High Resolution Instrument (HRI)
visible imager on the EPOXI \citep{epoxi} spacecraft when it was located
$\sim 0.1\,$AU from Earth. Observations with the EPOXI HRI were requested
in an attempt to constrain the microlensing parallax effect and obtain
precise mass measurements of the MOA-2009-BLG-266Lb planet and its
host star. We could obtain these observations because our target
field was able to provide a better test of newly installed pointing control
software than a less dense stellar field.

\begin{figure}
\centering
\includegraphics[width=16cm]{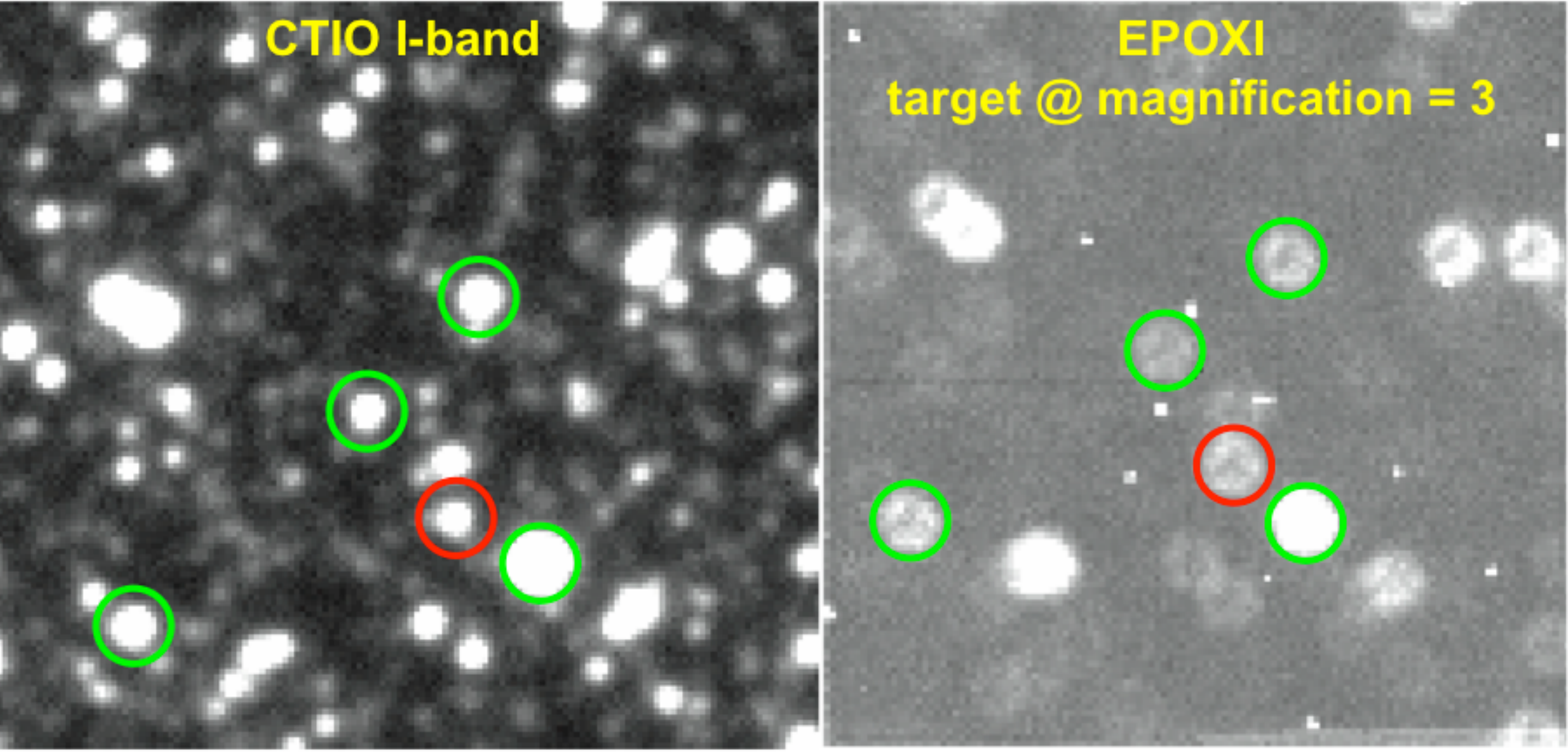}
\caption{\noindent Comparison of CTIO $I$ band (left) and EPOXI (right) images of the
MOA-2009-BLG-266 target, indicated by red circle. North is to the right and East is
up in the CTIO image, and the EPOXI image is rotated slightly from this orientation. 
The green circles indicate the 4 photometry comparison stars, which each have
$V-I$ within 0.9 mag of the target star. The comparison stars also have only one
star identified in the CTIO frames within 3 arc sec of the matched position of the
EPOXI star. 
}
\label{fig-ctio_epoxi}
\end{figure}

The EPOXI data consist of 4127 50 sec exposures
with the ``clear-6" filter. 
To minimize data transfer requirements, the data were 
taken as $128\times 128$ and $256\times 256$ pixel sub-frames. The
first 3375 exposures used $128\times 128$ sub-frames, and the last
752 images were $256\times 256$ sub-frames. However, the pointing stability
was such that the target occasionally drifted out of the $128\times 128$ sub-frames,
and it was only possible to do photometry on 2900 of these 3375 
$128\times 128$ pixel images. An example of one of these 
$128\times 128$ pixel images is shown on the right side of
Figure~\ref{fig-ctio_epoxi}.

The instrumental point-spread function (PSF)
of the High Resolution Instrument (HRI) on the
Deep Impact probe is strongly dependent on the color of the target
star.  In addition, the instrument is permanently defocussed, yielding
a toroidal--shaped PSF.  This is clearly not optimal for crowded field
photometry, which typically requires a spatially varying empirical PSF
model for deblending.  Since we performed photometry using the
Daophot/Allstar/Allframe software suite \citep{allframe}, we first
required a model of the instrumental PSF that is usable by Daophot.

Instrumental PSFs have been generated by \citet{barry_psf} for the HRI using the
Drizzle algorithm \citep{drizzle}.  In this process several hundred images
of a standard star were added together to create a ten--times
oversampled PSF model appropriate for that object.  To approximate the
PSF of MOA-2009-BLG-266, with $V - I = 1.82$, we coadded the instrumental PSFs of
GJ436 with $V - I = 2.44;$ and XO-2 with $V - I = 0.75$
with weightings of 0.795 and 0.205, respectively.  This composite PSF
was added to an otherwise empty image in a 3 by 3 grid, with each
realization downsampled to standard resolution using a center pixel
shifted by $\pm$ 5 pixels in x and/or y.  Daophot was then run on this
image, using all 9 images to build its own internal,
double--resolution PSF model using an empirical function plus ``lookup
table".

Approximately 1\% of pixels in our 50s observations contain signatures
of cosmic rays.  We filtered these pixels using an algorithm that
identifies features sharper than expected from the PSF, through
pairwise comparison of neighboring pixels.  These pixels were masked
and objects underneath these pixels ignored when generating
lightcurves.  We used Daophot to detect stars in each image, and then
Allstar to perform joint PSF photometry on all stars in a given image.
We generated a master starlist by matching the results of the Allstar
analysis using Daomatch and Daomaster.  This starlist was then sent to
Allframe, which simultaneously photometered all images in a
self--consistent manner with regards to centroiding and deblending.

To assemble the final light curves, we first aggregated the Allframe
measurements of a given star across all images.  
% To compensate for
% known variations in the pixel response across the CCD, we next
% computed a correction function to the measured brightnesses, assuming
% that the comparison stars are constant in brightness.  This magnitude
% correction was a linear function of the $x$ and $y$ position in the
% image, and generated using 83 comparison stars.  All measurements of
% all stars were then corrected using this function, yielding the final
% lightcurve of MOA-2009-BLG-266Lb, as well as the brightnesses of the comparison
% stars that were used to calibrate our photometry to the OGLE system.
Due to the difficulty of obtaining precise flat field images in space, we
cannot calibrate these images using the same methods as would be
used for ground-based images. As a result, the light curves of all the
stars observed by EPOXI/HRI show variations at the $\sim 1\,$\% level on
a time scale of a few hours, which is the time that it takes for the pointing
to drift a distance of order the PSF FWHM. Because this is the dominant 
term in the EPOXI/HRI photometry errors, we bin the data at an interval of
2.4 hours, which seems to remove most, but not all, of the correlations. 
This gives the light curve shown in the inset in the upper right
hand corner of Figure~\ref{fig-lc}. 

We had also hoped to get EPOXI/HRI observations after the
MOA-2009-BLG-266 microlensing event had returned to its
baseline brightness in March or April, 2010. Unfortunately, 
the EPOXI operations team was busy with preparations for the
November, 2010 encounter with comet Hartley 2, so no baseline
observations were possible. Therefore, we have determined
the baseline brightness in the EPOXI by comparing the EPOXI
images to $V$ and $I$ band CTIO images taken in June 2010, when
the microlensing event had returned to baseline. The $I$ band CTIO image
is compared to one of the EPOXI frames in Figure~\ref{fig-ctio_epoxi}.
Because of the relatively large EPOXI/HRI PSF, we consider only stars 
in the EPOXI images that have only one counterpart star within a radius
of $3^{\prime\prime}$ of the position of the EPOXI star. We also limit 
our consideration to stars within 0.9 mag of the $V-I$ color of the
microlensed target star. There are 4 stars that satisfy this condition and
appear in more than half of the images in which the target star appears.
These are the 4 stars indicated by the green circles in Figure~\ref{fig-ctio_epoxi}. We fit
the mean ``clear"-filter EPOXI magnitude, $C_{EPOXI}$ to the instrumental 
CTIO magnitudes from the SoDoPHOT reductions, and this yields the 
following linear relationship between the average EPOXI 
magnitudes and CTIO magnitudes,
\begin{equation}
C_{EPOXI} = 0.520 I_{CTIO} + 0.480 V_{CTIO} \ .
\label{eq-Uepoxi}
\end{equation}
The fit to the magnitude of these 4 comparison stars gives $\chi^2=0.22$ for
2 degrees of freedom if the uncertainty in the EPOXI magnitudes is 
assumed to be 0.004 mag. We use this formula to determine the 
baseline $C_{EPOXI} $ magnitude, and we add this to the light curve
as a final measurement with an assumed uncertainty of 0.01 mag.

We note that this calibration procedure for the EPOXI data is probably
more reasonably considered to be a procedure to calibrate the source star flux 
instead of the baseline brightness, which includes the brightness of any
unresolved stars blended with the source. But we chose to treat the
unmagnified flux estimate as an estimate of the baseline brightness as
this is more conservative.

\section{Source Star Properties and Einstein Radius}
\label{sec-src_prop}

Planetary microlensing events typically have caustic crossing
or cusp approach features that resolve the finite angular size
of the source star, and MOA-2009-BLG-266 is no exception as
it has clear caustic crossing features. The modeling of such features
constrains the source radius crossing time, $t_\star$, and this can be quite 
useful because $t_\star$ can be used to determine
the angular Einstein radius, $\theta_E = \theta_\star t_E/t_\star$,
as long as the source star angular radius, $\theta_\star$ can be
determined. In events
such as MOA-2009-BLG-266, with a strong microlensing parallax signal,
$\theta_E$ can be combined with the parallax measurement to yield the
lens system mass. Therefore, it is important to make an accurate determination
of the angular radius of the source star, $\theta_\star$.

\subsection{Source Star Colors and Extinction}
\label{sec-col_extinct}

\begin{figure}
\centering
\includegraphics[width=13cm]{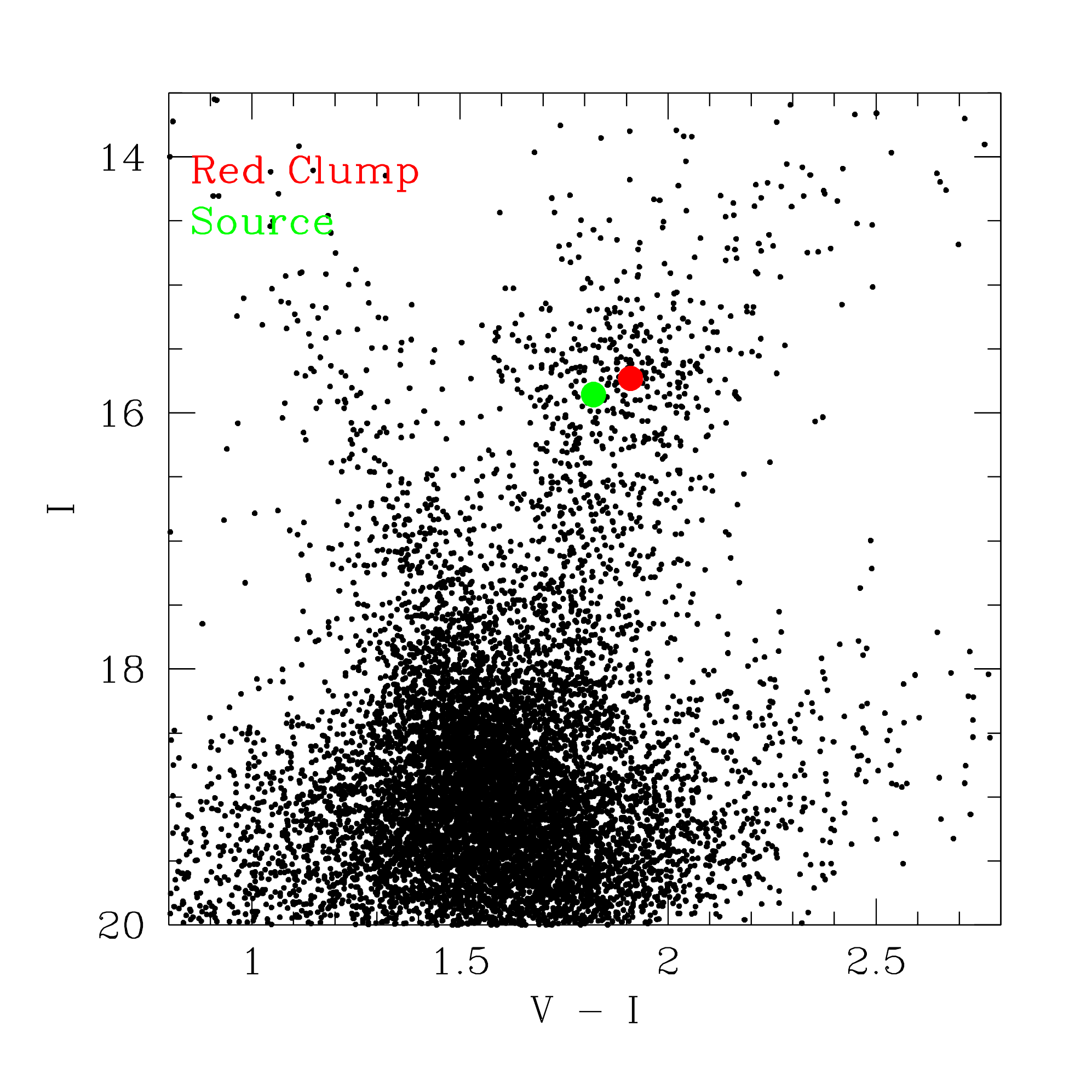}
\caption{\noindent Color-magnitude diagram of stars from the
OGLE-III database within $2'$ of 
the MOA-2009-BLG-266 source star. The center of
the red clump and the source star are indicated by
the red and green spots.
}
\label{fig-VIcmd}
\end{figure}

\begin{figure}
\centering
\includegraphics[width=13cm]{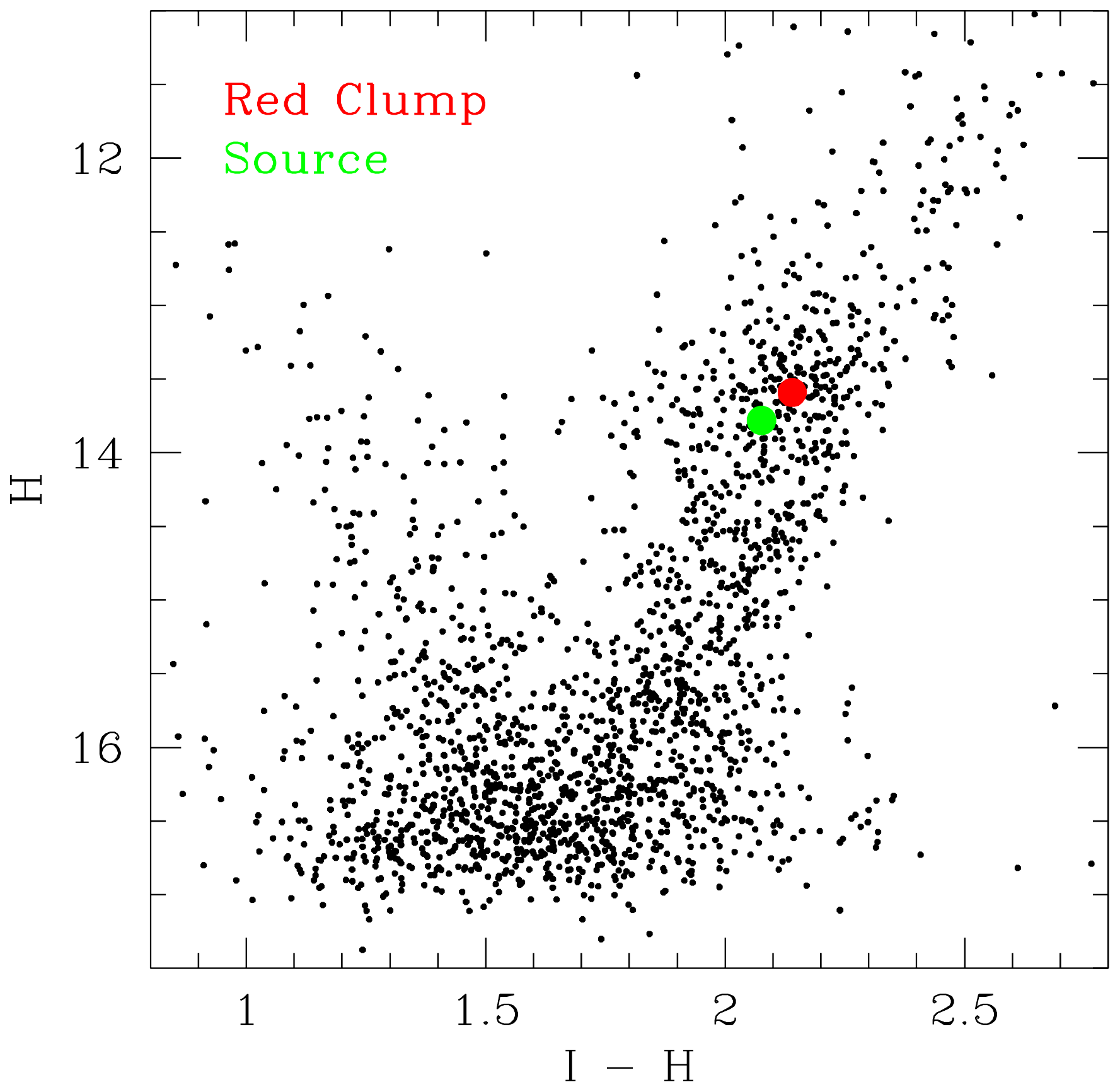}
\caption{\noindent The
$I-H$ color-magnitude diagram of stars within $2'$ of
the MOA-2009-BLG-266 source star, based on IRSF $H$-band
data calibrated to 2MASS and OGLE-III $I$-band photometry.
As in Figure~\ref{fig-VIcmd},
the red and green spots indicate the red clump center and
the source star.
}
\label{fig-IHcmd}
\end{figure}

The angular radius of the source star can be determined from its
brightness and color, once the effect of interstellar extinction has been
removed. We start from the CTIO $V$ and $I$ band magnitudes and the
IRSF $H$ band magnitude that
have been determined from the best fit model. The $V$ and $I$
band magnitudes have been calibrated to the 
OGLE-III system \citep{udalski08} and 
IRSF $H$ band has been calibrated to 2MASS
\citep{2mass_cal}\footnote{Improved calibrations are available at \\
{\tt http://www.ipac.caltech.edu/2mass/releases/allsky/doc/sec6\_4b.html}}.
The comparison between the 2MASS and the IRSF $H$-band data 
is subject to complications due to variability and
blending, because the 2MASS images,
with their $2"$ pixels, have significantly worse angular resolution than 
IRSF. This means that many of the apparent 2MASS ``stars" cannot 
be used for the calibrations because they are actually blended 
images of two or more stars that are resolved in the IRSF images.
This makes calibration of the CTIO $H$ band images difficult, because
of the relatively small $5.5\,{\rm arc\ min}^2$ CTIO $H$  field of view
(FOV). Fortunately, the IRSF FOV is $\sim 60\,{\rm arc\ min}^2$, and
it is possible to use over 400 unblended 2MASS stars for the $H$ band
calibration. These calibrations combined with the best fit light curve
models yield source magnitudes of
\begin{eqnarray}
\label{eq_s_magH}
H_s =& 13.780 \pm 0.030 \\
\label{eq_s_magI}
I_s =& 15.856 \pm 0.030 \\
V_s =& 17.677 \pm 0.030 \ ,
\label{eq_s_magV}
\end{eqnarray}
where the uncertainties are almost entirely due to the calibrations
(including the uncertainty in the OGLE-III calibration. These magnitudes
are indicated by the green dots in the color magnitude diagrams shown in
Figures~\ref{fig-VIcmd} and \ref{fig-IHcmd}.
The fit 
uncertainties are $\leq 0.005\,$mag in all three passbands, and the 
$u_0 > 0$ model predicts a source that is $0.005\,$mag brighter
than the best fit $u_0 < 0$ model.

\subsection{Source Star Radius}
\label{sec-src_rad}

We can use the magnitudes from equations 
\ref{eq_s_magH}-\ref{eq_s_magV} to determine the source star angular
radius, but first we must estimate the foreground extinction. We determine
the source radius using the method of Bennett et al.\ \citep{bennett-ogle109},
which is a generalization to three colors of an earlier two-color 
method \citep{albrow-97blg41}. Following this procedure, we find the
$VIH$ magnitudes of the center of the red clump giant distribution to be
\begin{eqnarray}
\label{eq_rc_magH}
H_{rc} =& 13.59 \pm 0.10 \\
\label{eq_rc_magI}
I_{rc} =& 15.73 \pm 0.10 \\
V_{rc} =& 17.64 \pm 0.10 \ ,
\label{eq_rc_magV}
\end{eqnarray}
for stars within $2'$ of the source star. These are indicated by the red spots
in Figures~\ref{fig-VIcmd} and \ref{fig-IHcmd}. Assuming a distance to
the source of $8.8\,$kpc \citep{rattenbury07}, we can use 
these red clump magnitudes to estimate the extinction,
which we find to be $A_H = 0.36$, $A_I = 1.22$, and $A_V = 2.14$,
following a $R_V = 2.77$ Cardelli et al.\ \citep{cardelli_ext} extinction law.
These then yield de-reddened magnitudes of 
$H_{s0} = 13.42$, $I_{s0} = 14.64$, and $V_{s0} = 15.54$.
Unlike the case for dwarf stars, the accuracy of the 
$V-I$, $V-H$, and $I-H$ surface
brightness-color  relations \citep{kervella04g} is similar, but the
$I-H$ relation yields an angular radius estimate that is
almost completely independent of the reddening law, if it follows
the Cardelli et al.\ extinction law \citep{cardelli_ext}, but this may be due to the fact that the
$A_I/A_H$ ratio doesn't vary much with this extinction law. In any
case, all three of these relations imply
that the angular radius of the 
source star is
\begin{equation}
\theta_\star = 5.2\pm 0.2\ \mu{\rm as}.
\label{eq2} 
\end{equation}
This and the source radius crossing time of
$t_\star = 0.326\pm 0.007\,$days imply that the
relative lens-source proper motion and Einstein radius are
\begin{equation}
\mu_{\rm rel} = \theta_\star/t_\star=5.86 \pm 0.26 \ {\rm mas}\ 
{\rm yr}^{-1},
\label{eq4}
\end{equation}
and
\begin{equation}
\theta_{\rm E} =\mu_{\rm rel}t_{\rm E} = 0.98\pm0.04\ {\rm mas},
\label{eq3}
\end{equation}
respectively.

\subsection{Limb Darkening}
\label{sec-limb}

During caustic crossings the lens effectively scans the source 
star with high angular resolution. As a result, the shape of the light 
curve reflects the underlying limb darkening of the 
star \citep{witt95,bennett96}. Hence, in order to analyze caustic-crossing 
events such as MOA-2009-BLG-266, one needs to account for the limb 
darkening appropriately. For the previous planetary microlensing 
events \citep{bond04,udalski05,ogle390,gould06,gaudi-ogle109,bennett08,
dong09,sumi10,janczak10,miyake11,batista11}, limb darkening has generally been 
treated within the linear limb-darkening approximation. In some cases, the 
two-parameter square-root limb-darkening law was used \citep{dong09,janczak10}, 
even though there has been no indication that the details of the limb-darkening 
treatment had any noticeable effect on the other model parameters
for a planetary microlensing event.

In the case of MOA-2009-BLG-266, there was reason to suspect that the treatment 
of limb darkening could be important. This is because, as we discuss below 
in Section~\ref{sec-mod}, there are two approximately degenerate microlensing parallax 
models that have slightly different binary lens parameters. The source crosses 
the caustics at a slightly different angle for the two models. This suggests 
that the detailed treatment of limb darkening might have some influence on the 
difference in $\chi^2$ between these two degenerate models. As shown by 
\citet{heyrovsky07}, using linear limb darkening may introduce 
photometric errors at the level of 0.01 due to the approximation itself and
the choice of method used for computing the linear model coefficients. 
In order to avoid introducing any such inaccuracies in the analysis of the event, 
we directly use the limb-darkening profile from a theoretical model atmosphere 
of the source star, instead of its analytical approximations.

Based on the location of the source star on the color magnitude diagram, 
we assume a temperature of $T_{\rm eff} \approx 4750\,$K, surface gravity 
of $\log g=2.5$, and solar metallicity. We use a model atmosphere from 
Kurucz's ATLAS9 grid
\citep{kurucz,kurucz93a,kurucz93b}\footnote{Updated online at http://kurucz.harvard.edu}
corresponding to these parameters. The raw model data provide values 
of the specific intensity as a function of wavelength for 17 different positions 
on the stellar disk. In order to obtain the light-curve-specific limb-darkening profile, 
we integrate the specific intensity over the relevant filter passband, weighted by 
the filter transmission, the quantum efficiency of the CCD, and interstellar 
extinction (see Section~\ref{sec-col_extinct}). In order to compute the limb darkening at 
an arbitrary position on the stellar disk, we interpolate the obtained points 
using cubic splines with natural boundary conditions \citep{heyrovsky03,heyrovsky07}. 
The light curve modeling code uses pre-computed tabulated intensity 
values for a sufficiently dense spacing of radial positions on the stellar disk.

This approach avoids a potential source of low-level systematic error 
without any degrees of freedom to the model. In Table~\ref{tab-chi2} we 
compare the results of our analysis with those obtained by the usual 
approach, using linear limb-darkening coefficients from Claret \citep{claret00}. 
For the parameters of the source star Claret \citep{claret00} provides 
coefficient values $u_\lambda=0.7844$, 0.7035, 0.6087, 0.4868, 0.4181, 
and 0.358 for the $V$, $R$, $I$, $J$, $H$, and $K_s$ passbands, 
respectively. Ttabulated intensity values give a $\chi^2$ improvement 
over the linear approximation of $\Delta\chi^2 = 7.27$ for the best-fit static 
models without orbital motion.   
So, the tabulated limb darkening tables fit the data somewhat better,
at least for the static lens case, but the implied planetary parameters
do not change significantly.

\section{Planet Characterization and Modeling}
\label{sec-pl_char_mod}

\subsection{Modeling}
\label{sec-mod}

The basic parameters for planetary events like MOA-2009-BLG-266
are straightforward to determine, as a reasonable estimate
can be made from the single lens parameters (found from a
fit with the planetary 
deviation excluded) and inspection of the light curve
\citep{gouldloeb92}. The main feature of the deviation is the half-magnitude
decrease in magnification centered at ${\rm HJD}' = 5086.5$. This
indicates that a planet is perturbing the minor (saddle) image created by the 
stellar lens and that the star-planet separation is
less than the Einstein radius. Such a light curve cannot be 
mimicked by non-planetary perturbations \citep{gaudi97}. The basic
planetary parameters can then be estimated following the arguments 
given in  \citet{sumi10}.
In practice, this is not how the parameters were determined, however.

We model the data using standard methods \citep{bennett-himag,dong06}
to extract the
precise parameters and uncertainties of the light curve fit. 
It is convenient to describe microlensing events in terms of the
Einstein ring radius, $R_E = \sqrt{(4GM_L/c^2)D_Sx(1-x)}$, 
which is the radius of ring image
seen when the source and (single) lens are in perfect alignment.
Here $M_L$ is the lens system mass, $x = D_L/D_S$, and 
$D_L$ and $D_S$ are the lens and source distances. 
Microlensing by a single lens, such as an isolated
star, is described by three parameters: the Einstein
radius crossing time, $t_E$, and the time, $t_0$, and distance 
(with respect to $R_E$),
$u_0$, of closest alignment between the source and lens
center of mass. 
Planetary microlensing events require three additional parameters: the
planet:star mass ratio, $q$, the star-planet separation, $s$, in units of $R_E$, 
and angle of the source trajectory with respect to the star-planet 
axis, $\theta$. The source radius crossing time, $t_\star$, is
also required because, like most planetary events, 
MOA-2009-BLG-266 has sharp light curve
features that resolve the angular size of the source star.

The MOA and Canopus data for the event were modeled immediately
upon the detection of the planetary perturbation using the method of 
\citet{bennett-himag}, supplemented with the addition of the hexadecapole
approximation \citep{pej_hey,gould-hex}. 
This found the basic solution that we present here,
plus a disfavored alternative $s>1$ solution, which was excluded within
hours when the planetary deviation data from South Africa, Israel, and
Chile became available. The $s<1$ solution was refined as more
data came in, and the two degenerate solutions that we present here
emerged when microlensing parallax was added to the modeling.

We also conducted a blind search of parameter space using 
 the approach of \citet{dong06}, where the binary parameters 
$s$, $q$, and $\theta$ are fixed at a grid of values, while the 
remaining parameters are allowed to vary so that the model light 
curve results in minimum $\chi^2$ at each grid point.  A 
Markov Chain Monte Carlo method was used
for $\chi^2$ minimization.  Then, 
the best-fit model is obtained by comparing the $\chi^2$ minima 
of the individual grid points.

\subsection{Best-fit Model}
\label{sec-best_mod}

The modeling indicates that the perturbation of MOA-2009-BLG-266 is produced 
by the crossing of a clump giant source star over the planetary 
caustic produced by a low-mass planet. As we discuss below
in Section~\ref{sec-orb}, the orbital motion of the planet is not needed to
describe the light curve. Assuming a static lens sytem,
the best fit values of 
the planet/star mass ratio and normalized star-planet separation are
$q = 5.425\times 10^{-5}$ and $s = 0.91422$,
respectively.
The values of other lensing parameters are listed in 
Table~\ref{tab-fitpars}. This table also lists the best fit parameters for 
fits including orbital motion.  The inclusion of orbital motion improves
$\chi^2$ by $\Delta\chi^2 = 3.24$ for 2 fewer degrees of freedom,
but it also changes the planet/star mass ratio and normalized 
star-planet separation to $q = 5.815\times 10^{-5}$ and $s = 0.91429$.
As discussed below in Section~\ref{sec-par}, the inclusion of microlensing
parallax adds a parameter degeneracy that takes $u_0 \rightarrow -u_0$
and $\theta \rightarrow -\theta$, which corresponds to a reflection of the
lens plane with respect to the geometry of the Earth's orbit. We
find that the model with $u_0 > 0$ is favored by $\Delta\chi^2 = 13.39$
for static models and $\Delta\chi^2 = 6.31$ for models with orbital 
motion.
The $\chi^2$ contribution of the individual data sets for the
best models with and without orbital motion is shown in Table~\ref{tab-chi2}.
We note that this mass ratio is 
the lowest of planets yet to be discovered by the microlensing method.

\begin{figure}
\centering
\includegraphics[width=15cm]{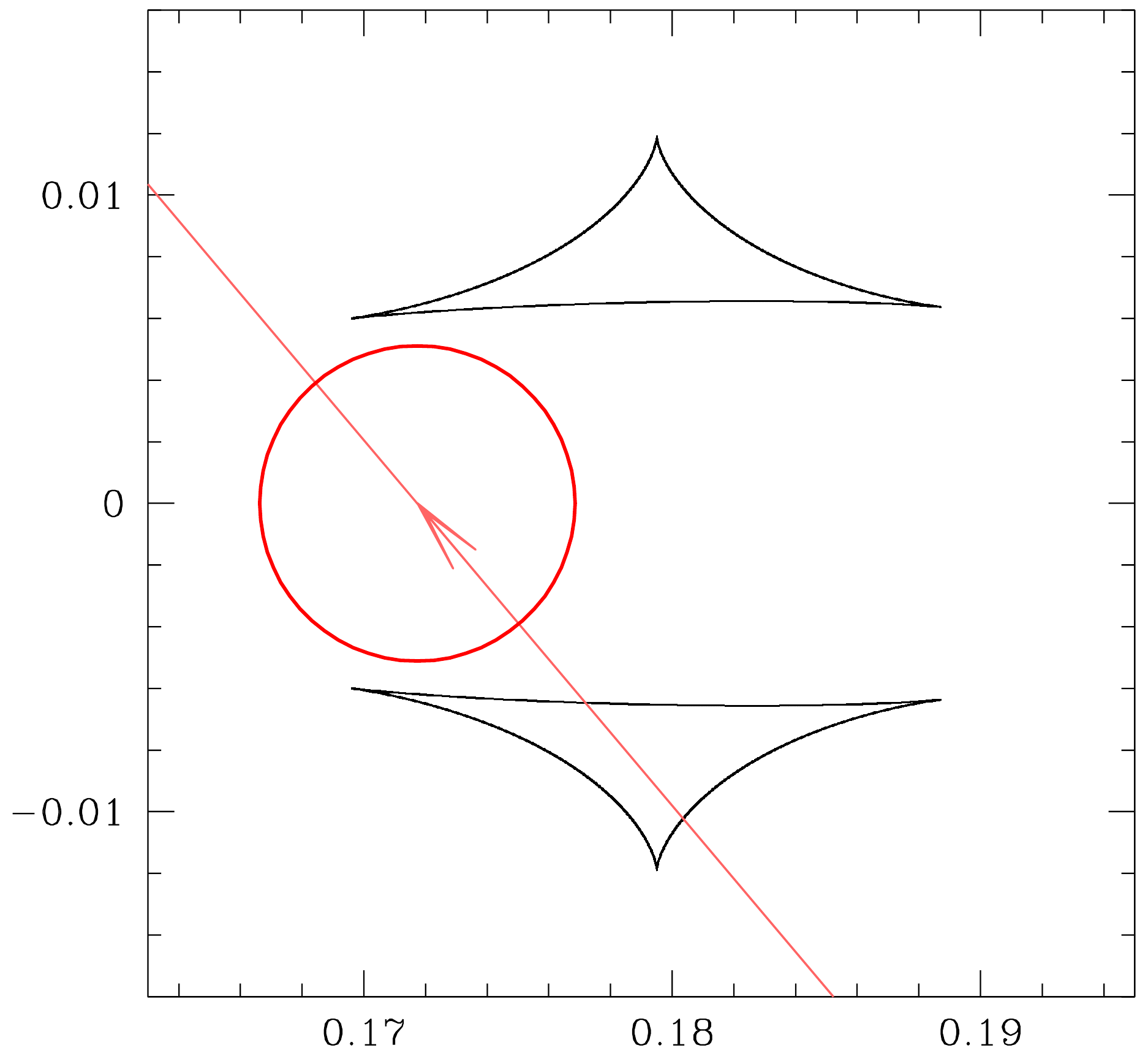}
\caption{\noindent The planetary caustic for the best fit
MOA-2009-BLG-266 light curve model is shown in black.
The red circle indicates the size of the source star and the red
line indicates its trajectory. The coordinates are in Einstein radius
units with the center of mass at the origin.
\label{fig-caustic}
}
\end{figure}

Figure~\ref{fig-lc} shows the best-fit model curve compared to the light curve
data.  Figure~\ref{fig-caustic} compares the planetary caustic
geometry to the source size and trajectory. The two triangular
shaped caustics indicate a minor image caustic perturbation, which
is seen when the star-planet separation is less than the Einstein
radius ($s < 1$). The strongest feature in such a minor image
caustic crossing event is the large decrease in magnification
at ${\rm HJD}\sim 2455086.5$
when the source is between the caustics and the minor image
is essentially destroyed. This feature is surrounded by two positive
light curve bumps caused by the source passing over a caustic
or passing in front of the cusps. There are no known
non-planetary light curve perturbations that can produce such
a feature in the light curve \citep{gaudi97}.
For MOA-2009-BLG-266, the local
light curve minimum between the caustic crossings has a short
duration of $\sim 3.7\,$hours, which is much smaller than the 
caustic crossing durations of $> 20\,$hours. This is due to the
fact that separation between the two triangular minor image
caustics is only slightly larger than the diameter of the source star.

\subsection{Orbital Motion}
\label{sec-orb}

Like most low-mass planetary microlensing events, MOA-2009-BLG-266
can be well modeled without including any orbital motion of the
planet about the host star. But, while we haven't measured the
light curve precisely enough to measure orbital motion parameters,
the planetary orbital motion does influence the precision to which
the parameters can be measured from the light curve. In particular,
orbital motion allows the planetary caustic to move with respect to
the center of mass of the system. Thus, the planetary caustic
can be either larger or smaller than the value determined from static
lens models, so the mass ratio is not measured as precisely as
the static models would imply. In addition, the source radius crossing time
is also determined by the time it takes the source to cross the
sharp light curve features of the planetary caustic, so this also
depends on the orbital motion of the planet and is less precisely
determined than the static models would imply. 

We include orbital motion using the parameterization used for the
analysis of the two-planet event 
OGLE-2006-BLG-109Lb,c \citep{bennett-ogle109}, with the $x$-axis defined by
the vector separating the star and planet at
${\rm HJD} = 2455086$. The main orbital motion 
parameters are $\dot s_x$ and $\dot s_y$, which
describe the instantaneous planet velocity at the time ${\rm HJD} = 2455086$.
This parameterization also includes the orbital period, $T_{\rm orb}$,
but this parameter has a very small effect on the $\chi^2$ value if 
it is in the range of physically reasonable values. So, for many of
our calculations, we have left $T_{\rm orb}$ fixed at a physically
reasonable value.
Independent calculations with a slightly different orbital motion
parameterization \citep{skowron11} reached identical conclusions.

The effect of the orbital motion on the other light curve 
parameters can be seen
in Table~\ref{tab-fitpars}, which shows the parameters and error bars
for models with and without orbital motion. The inclusion of orbital
motion shifts the values of $q$ and $t_\star$ significantly, by
2.6 and 3.8 times the error bars of the static solution, respectively.
Orbital motion also increases the error bar on $q$ by a factor of 3.6 and
the error bar on $t_\star$ by a factor of 5.7, but the error bars
on the other parameters don't change significantly, except for
the error bar on $t_0$, which grows by a factor of 1.6. The 
error bars on the implied physical parameters, shown in 
Table~\ref{tab-pparam}, also don't change very much when 
orbital motion is included. However, the central values of
the physical parameters do change by as much as $0.4\sigma$.

While the light curve has not been measured precisely enough
to significantly constrain the orbital motion, the orbital motion can
be constrained with the requirement that the planet
be bound to its host star. Such a constraint requires that the
mass and distance of the host star be known, but the light curve
does provide this information as shown below in eqs.~\ref{eq-M} and 
\ref{eq-Dl}. The light curve parameters include the transverse
host star-planet separation, $s$, and the transverse components
of the planet velocity, $\dot s_x$ and $\dot s_y$. One option
is to enforce a model constraint that the orbital motion parameters
are consistent with a physical circular orbit. This is equivalent to
imposing a constraint on the distance to the 
source \citep{bennett-ogle109}, which we take to be
$D_S = 8.8\pm 1.2\,$kpc based on the Galactic 
longitude of this event \citep{rattenbury07}. This constraint
has been employed for the best fit model shown in
Table~\ref{tab-fitpars}. But, such a constraint is inconvenient to use in
our Markov Chain Monte Carlo (MCMC) calculations to determine 
the distribution of allowed light curve and physical parameters,
as it makes it more difficult to obtain well sampled Markov Chains.

A slightly weaker, but more general, constraint can be obtained by requiring
that the orbital velocity not be too large to allow the planet to be
gravitationally bound to the star, since the probability of lensing by
a planet not bound to the lens star is $\simlt 10^{-8}$. The transverse
velocity components allow us to compute lower limits on the 
planetary kinetic energy and gravitational binding energy (or
an upper limit on the absolute value of the binding evergy).
The requirement that the total
energy $< 0$ yields an upper limit on the transverse
planet velocity \citep{dong09}:
\begin{equation}
\dot s_x^2 + \dot s_y^2 \leq {2GM_L\over d_\perp R_E^2} 
                                   = {2GM_L\over s (\theta_E D_L)^3} \ ,
\label{eq-Econst}
\end{equation}
where $d_\perp$ is the transverse star-planet separation.
The $R_E^3 = (\theta_E D_L)^3$ factor in the denominator is needed because the
planet-star separation, $s$, and transverse velocity components
use the Einstein radius as their unit of length. 

The constraint, eq.~\ref{eq-Econst}, has been used in all of our MCMC
calculations to determine the allowed parameter distributions, and we have
also added a $\Delta\chi^2 = 4$ penalty to potential MCMC links with a
$\dot s_x^2 + \dot s_y^2$ of more than half the upper limit in eq.~\ref{eq-Econst}.
Such parameter sets are unlikely because they require a kinetic energy
higher than the average value and small values for the separation and velocity
along the line of sight. Attempts to find best solutions with the 
eq.~\ref{eq-Econst} in place of the circular orbit, source distance constraint
did not yield better solutions than the one shown in Table~\ref{tab-fitpars}.

\subsection{The Parallax Effect}
\label{sec-par}

The microlens parallax is defined by the ratio of the Earth's 
orbit to the physical Einstein radius projected on the observer 
plane, $\tilde{r}_{\rm E}$, i.e.,
\begin{equation}
\pi_{\rm E}={{\rm AU}\over \tilde{r}_{\rm E}}.
\label{eq6}
\end{equation}
Lens parallaxes are usually measured from the deviation of the light curve 
from those of standard (single or binary) lensing events due to the 
deviation of the source trajectory from a straight line caused by 
the orbital motion of the Earth around the Sun 
\citep{gould-par1,macho-par1}. But it is also possible to detect microlensing
parallax using observations from a spacecraft in a heliocentric orbit
\citep{refsdal-par,dong-ogle05smc1}, and such satellite observations have
the potential to significantly increase the number of events for which the
microlensing parallax effect may be detected.

For the event MOA-2009-BLG-266, the parallax effect is firmly 
detected.  We find that the $\chi^2$ difference between the 
(static) best-fit models with and without the parallax effect is $\Delta\chi^2
=2789.3$, which implies that microlensing parallax is detected
at the $\sim 53 \sigma$ level.  The difference 
between the parallax and non-parallax models can be seen in Figure~\ref{fig-lc},
where the best fit model is plotted as a solid black curve and the best fit
non-parallax model is the grey dashed curve. Most of the parallax signal
comes from the data outside of the planetary deviation. The light curve
without parallax lies below the observed data prior to the planetary 
perturbation and above the data after the light curve peak. Most
of the signal comes from the MOA data, but good coverage of the
global light curve shape from CTIO and Canopus has enabled these
light curves to contribute to the parallax signal.

There are a number of degeneracies that often affect the parallax
parameters of an event.
For events with parallax effect, a pair of source 
trajectories with the impact parameter and source trajectory angle 
of $(u_0,\theta)$ and $(-u_0, -\theta)$ can yield degenerate 
solutions \citep{smith03}.  Without parallax, this transformation
is a trivial redefinition of parameters, but with parallax, we have the
reference frame of the Earth's orbit, which allows us to distinguish
between two solutions that differ by a reflection of the lens plane. 
For single lens events with $t_E$ as
small as $\sim 60\,$days, there is usually an additional degeneracy
known as the jerk-parallax degeneracy \citep{gould-jerk}, but the
additional light curve structure due to the planet removes some of the
parameter degeneracy. As a result, the  
$(u_0,\theta) \leftrightarrow (-u_0, -\theta)$ degeneracy and the jerk-parallax
degeneracy are replaced by a single parameter degeneracy that makes
the $(u_0,\theta) \leftrightarrow (-u_0, -\theta)$ and changes the parallax
parameter, $\piEbold$. This degeneracy yields the two solutions
with similar parameters, but when orbital motion of the planet is
ignored, there are significant differences in the other model 
parameters besides $u_0$ and $\theta$. With no orbital motion,
the $u_0 > 0$ solution is favored by
$\Delta\chi^2 = 13.39$. Formally, this is quite significant as
the $u_0 < 0$ solution would be disfavored by a formal probability of 
$e^{-\Delta\chi^2/2} \approx 0.0012$, but we must also consider
possible systematic errors that may influence the
$\Delta\chi^2$ value between the $u_0 > 0$ and $u_0< 0$ solutions,
as well as the orbital motion of the planet.

This concern about possible systematic errors was the reason for
the careful limb darkening treatment described in Section~\ref{sec-limb}.
In addition, we also considered a number of different photometric
reductions of the data sets that contribute the most to the detection
of the parallax signal. These were the MOA, Canopus, and CTIO $I$ and $H$
band data sets. With these different photometric reductions, we found
that the $u_0 > 0$ solution was always favored by a similar $\Delta\chi^2$ 
difference, although the SoDoPHOT reduction of the CTIO $I$ band data
set and the DoPHOT reduction of the CTIO $H$ band data set favored
the $u_0 < 0$ solution by a somewhat larger $\Delta\chi^2$ difference.
Table~\ref{tab-chi2} indicates that the $\chi^2$ difference between the
$u_0 > 0$ and $u_0 < 0$ solutions is spread over a number of different
data sets.

The inclusion of the orbital motion parameters discussed in 
Section~\ref{sec-limb} has a significant effect on the 
$(u_0,\theta) \leftrightarrow (-u_0, -\theta)$ degeneracy. When the
orbital motion parameters are included, the best fit ($u_0 > 0$)
model improves its $\chi^2$ value by $\Delta\chi^2 = 3.24$, but the
$\chi^2$ improvement for the $u_0 < 0$ models is even
greater, $\Delta\chi^2 = 10.34$, so that the $\chi^2$ difference 
between the $u_0 < 0$ and $u_0 > 0$ solutions drops to 
$\Delta\chi^2 = 6.29$ when the planetary orbital motion parameters
are included in the models. But, an even more significant difference
is that the differences between the other parameters for these
degenerate models largely disappear when orbital motion is included.
The added degrees of freedom provided by the orbital motion 
parameters appear to be larger than the light curve difference 
enforced by the $(u_0,\theta) \leftrightarrow (-u_0, -\theta)$
transformation. As a result, once orbital motion is included,
the $(u_0,\theta) \leftrightarrow (-u_0, -\theta)$ degeneracy has
no obvious effect on the determination of the physical parameters
of the event.

\begin{figure}
\centering
\includegraphics[width=10cm]{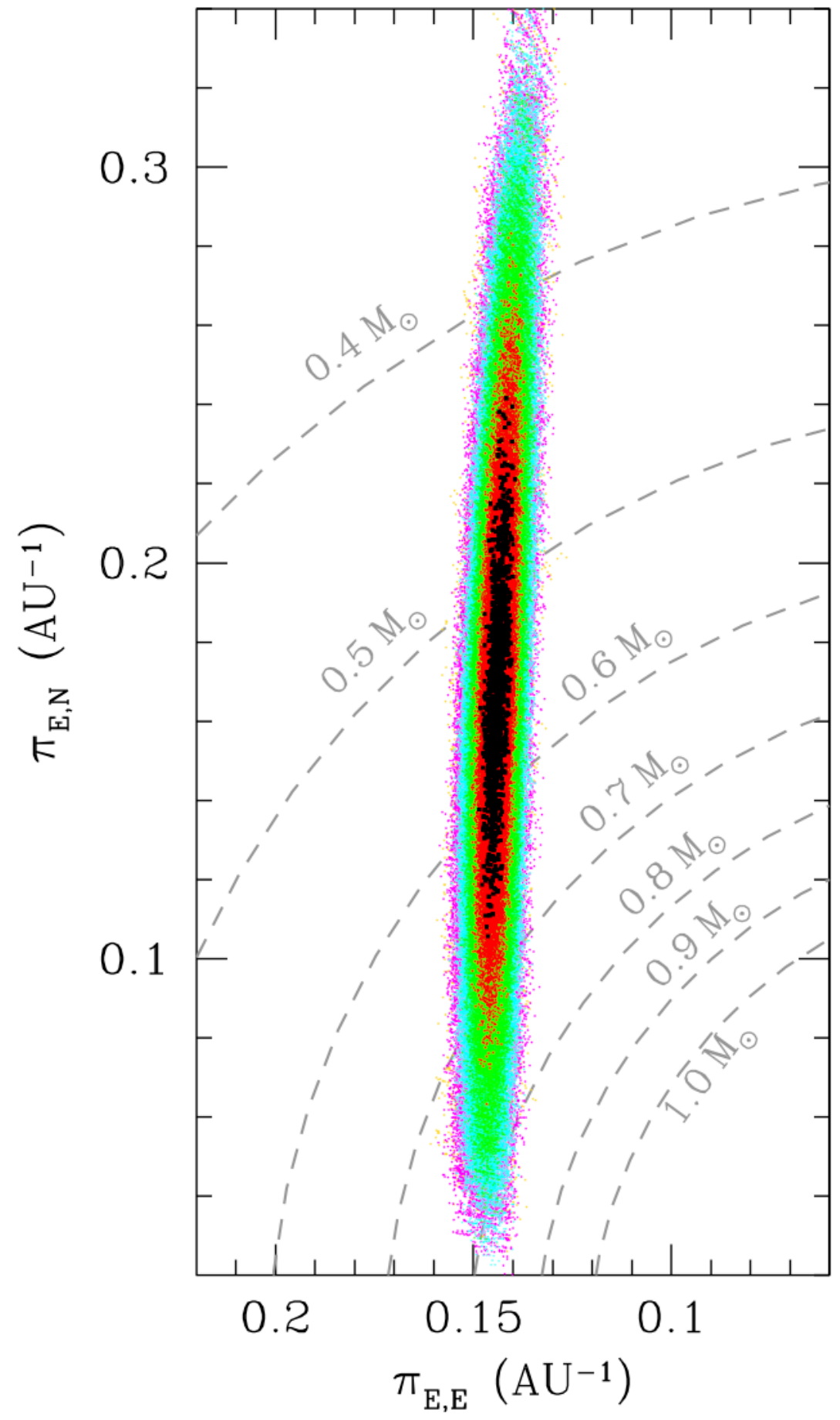}
\caption{\noindent 
Distribution of the microlensing parallax vector 
$\piEbold = (\pi_{{\rm E},N},\pi_{{\rm E},E})$ 
values that are consistent with the
observed light curve taken from our MCMC runs in the regions of
the two local $\chi^2$ minima with $u_0 < 0$ and $u_0 > 0$. The
points are color coded based on their difference from the
$\chi^2$ minimum of 6050.2, with points with $\Delta\chi^2 \leq 1$, 4,
9, 16, 25, 36, and $\Delta\chi^2 > 36$ represented by black, red, green, cyan, 
magenta, and gold
colored dots, respectively. The gray dashed circles indicate
contours of constant $\pi_E = 1/\rep$ and therefore
(approximately) constant mass. These assume the mean angular Einstein radius
from our MCMC calculations,
$\theta_E = 0.985\,$mas.
\label{fig-piE}
}
\end{figure}

Figure~\ref{fig-piE} shows the distribution of parallax parameters,
$(\pi_{{\rm E},N},\pi_{{\rm E},E})$ or equivalently, ($\pi_E,  \phi_E$)
found by our MCMC simulations. This plot includes 11 separate
MCMC chains with a total of 593,000 links as discussed in 
Section~\ref{sec-uncert}. The distributions
for both solutions are highly elongated along the 
$\pi_{{\rm E},N}$ axis. This is due to the fact that the Earth's acceleration
is almost entirely in the east-west direction, when projected on the
plane perpendicular to the line of sight to the Galactic bulge.
Figure~\ref{fig-piE} also shows contours of constant $\pi_E$, which
are labeled by the (approximate) corresponding
lens mass. The lens
mass depends on the angular Einstein radius, which our MCMC calculations
determine to be $\theta_E = 0.98\pm 0.04\,$mas. However, these mass
contours in Figure~\ref{fig-piE} are only approximate because they do not include any
correlations between the $\pi_E$ and $\theta_E$ values. These correlations
are properly incorporated into our MCMC calculations, which yield the
host star and planet masses, $M_\star = 0.56\pm 0.09\msun$ and 
$m_p =  10.4 \pm 1.7\mearth$, located at a distance of $D_L = 3.04\pm 0.33\,$kpc.
Assuming a random orientation of the orbit, we estimate a semi-major axis of
$a = 3.2{+1.9\atop -0.5}\,$AU. If we assume a standard position for the
snow line, $\sim 2.7 (M/M_\odot)\ {\rm AU}$ \citep{kennedy_snowline}, then
the planet orbits at about twice the distance of the snow line, similar to 
the position of Jupiter in our own solar system. Thus, the planet
might be considered to be a "failed Jupiter" core as predicted by the
core accretion theory \citep{thommes08,ida05}, in which the rock-ice core only
reaches $\sim 10 \mearth$ after the hydrogen and helium gas 
in the proto-planetary disk has dissipated.

In principle, any orbital parallax signal can be mimicked by so-called 
xallarap, i.e., orbital motion of the source about a companion 
\citep{griest92,han97}.  However, this would require very special 
orbital parameters, basically mimicking those of the Earth 
\citep{smith02}.  We search for such xallarap solutions over 
circular orbits, i.e., with 3 additional free parameters (orbital phase, 
inclination, and period).  We find a $\chi^2$ improvement of 3.9
relative to the parallax solution for 3 degrees of freedom, or 
3.4 with the period fixed at $P=1$ year (2 degrees of freedom).  
These improvements have no statistical significance and are to 
be compared with the improvement of $\Delta\chi^2=2789.3$ for the 
parallax solution relative to the no-parallax solution.  Therefore, 
we conclude that the light curve distortions are due to parallax 
rather than xallarap \citep{poindexter05}.

Our final fit parameters are listed in Table~\ref{tab-fitpars}.
The parameters  $\pi_{E,N}$ and
$\pi_{E,E}$ are the North and East components of the microlensing parallax
vector, $\piEbold$. The uncertainty 
in $\pi_{E,N}$ is an order of magnitude larger than
the uncertainty in $\pi_{E,E}$ because the
projected acceleration of the Earth is largely in the East-West
direction during this event.

\subsection{Parameter Uncertainties}
\label{sec-uncert}

Uncertainties in the parameters have been determined by a
set of 11 Markov Chain Monte Carlo (MCMC) runs with a 
total of approximately 593,000 links. Eight of these MCMC
runs have been in the vicinity of the $u_0 > 0$ solution and
the other three have been in the vicinity of the $u_0 < 0$ local $\chi^2$
minimum. Due to the $\chi^2$ difference,
$\Delta\chi^2 = 6.29$ between these
local $\chi^2$ minima, we include a weight factor to our
sums over the Markov chains so that the disfavored
$u_0 < 0$ solutions are disfavored by an amount
appropriate for their $\chi$ difference, $e^{-\Delta\chi^2} = 0.043$.
The mean parameter values for these solutions 
and their uncertainties are shown in
Table~\ref{tab-fitpars}. For most parameters, these are given
by the weighted averages over the 11 Markov chains, but for
$u_0$ and $\theta$, we have included only the 8 Markov
chains with $u_0 > 0$ and $\theta > 0$. Due to the large
difference in these parameters in the vicinity of the two solutions,
the mean values would be values that are inconsistent
with both solutions if we had used both solutions for these sums.
For the remaining parameters, except $\dot s_x$ and $\dot s_y$,
the parameter distributions for the vicinities of the two
local $\chi^2$ minima are nearly identical.

\subsection{Physical Parameters}
\label{sec-phys_par}

The source radius crossing time, $t_\star$, is an important parameter
because it helps to determine the angular Einstein
radius, $\theta_E = \theta_\star t_E/t_\star$, as discussed in 
Section~\ref{sec-src_rad}. When this is combined with
the measurement
of the microlensing parallax signal, we are able to determine the
mass of the lens system  \citep{gould-par1}, 
\begin{equation}
M_L = {\theta_E c^2 {\rm AU}\over 4G \pi_E}
    = {\theta_E \over (8.1439\,{\rm mas})\pi_E} \msun 
    \approx 0.57\msun \ .
\label{eq-M}
\end{equation}
if we assume that the favored parameters of the best fit
($u_0 > 0$) solution are correct.
The lens system distance can also be determined
\begin{equation}
D_L={{\rm AU}\over \pi_{\rm E}\theta_{\rm E}+\pi_S}
\approx 3.2 \ {\rm kpc},
\label{eq-Dl}
\end{equation}
assuming that the distance to the source, $D_S = 1/\pi_S$, 
is known. Note that these values from best fit solution
are not identical to the central values from our full MCMC analysis,
although they are very close.

In order to determine the physical parameters of this
planetary lens system, it is important to include the effects of
correlations of the parameters and the uncertainties in
external measurements, such as the determination of 
$\theta_\star$. Such a calculation is easily done with 
MCMC simulations. As discussed in Sections~\ref{sec-par} and \ref{sec-uncert}, we
have run 11 MCMCs in the vicinity of both the 
degenerate $u_0 > 0$ and $u_0 < 0$. The distribution
of the parallax parameters for these solutions is shown
in Figure~\ref{fig-piE}. The gray, dashed circles in this figure
show the contours of constant $\pi_E$, which correspond to 
contours of constant mass by equation~\ref{eq-M}. However, this
correspondence is only approximate because $\theta_E$ is
slightly correlated with $\pi_E$.

These MCMC simulations can also be used to determine the physical
parameters of the host star and its planet, which are summarized 
in Table~\ref{tab-pparam}. This is essentially a Bayesian analysis, but the
only non-trivial prior that we impose is the assumption that the orbital 
orientation is random, which is used to estimate the semimajor axis,
$a$, based on the measured two-dimensional separation in the plane
of the sky. If planets are much more common at very small or very large
separations, then the planetary detection would imply a bias that 
violates this assumption. But, the available evidence indicates that 
planet frequency does not have a sharp dependence on the 
semimajor axis, so this assumption seems reasonable. Our MCMC
calculations assumed a fixed distance of $8.8\,$kpc to the source star,
due to its position on the more distant end of the Galactic bar. 
We have adjusted the uncertainties in Table~\ref{tab-pparam} to 
include the 5\% spread in the distance to bulge clump stars measured
in this direction \citep{rattenbury07} (although the effect is quite small.)
The probability distributions for the host star and planet masses and distance
($M_\star$, $m_p$ and $D_L$)
are nearly Gaussian, so they are well described simply by their mean
values and dispersions. This is not the case for our estimate of the semi-major
axis, $a$, which has a $2\sigma$ (95\% confidence level) range of 
2.3-$12.9\,$AU.

We therefore conclude that the MOA-2009-BLG-266Lb the planet is a
$\sim 10\mearth$ planet at a separation of $\sim 3\,$AU. 
In the core accretion model of planet
formation, the snow line is an important location where the density
of solid material jumps by about a factor of five due to the 
condensation of ices (mostly water ice)
 \citep{ida05,lecar_snowline,kennedy-searth,kennedy_snowline,thommes08}.
Assuming a standard position for the
snow line, $\sim 2.7 (M/M_\odot)\ {\rm AU}$, we find that 
the planet is located at about twice the distance of 
the snow line, similar to 
the position of Jupiter in our own solar system. It is therefore a prime
candidate to be a ``failed Jupiter core\rlap," which grew 
by the accumulation of solid material to $\sim 10\mearth$, but was unable to 
grow into a gas giant by the accretion of Hydrogen and Helium because
the proto-planetary disk had lost its gas before the planetary core was massive 
enough to accrete it efficiently.

The mass measurements of the planet and host star given in 
Table~\ref{tab-pparam} have uncertainties of about 16\%, which
is dominated by the uncertainty in $\pi_{E,N}$. This uncertainty
can be reduced 5-10 year hence when the source and planetary 
host stars have separated enough to allow their relative
proper motion to be measured \citep{bennett07}. Since $\piEbold$
is parallel to the lens-source proper motion, this will reduce the
uncertainty in $\pi_E$ to a value much closer to the 2.4\% uncertainty
in $\pi_{E,E}$, which should reduce the uncertainties in the star
and planet masses to $< 5$\%. Our existing VLT/NACO
observations indicate that the combined $H$-band flux of the source and 
host star is $H = 13.77\pm 0.05$, which is consistent with our prediction
that the host star should be $\sim 75$ times fainter than the source and
indicate no neighbor stars that might interfere with the detection of the
source-host star relative motion. So, we expect that it will be
feasible to improve these mass measurements in the future.

We find a host star of mass $M_\star = 0.56\pm 0.09\msun$ orbited by a planet of mass 
$m_p =  10.4 \pm 1.7\mearth$, located at a distance of $D_L = 3.04\pm 0.33\,$kpc.
Assuming a random orientation of the orbit, we estimate a semi-major axis of
$a = 3.2{+1.9\atop -0.5}\,$AU and an orbital period of
$P = 7.6{+7.7\atop -1.4}\,$yr. If we assume a standard position for the
snow line, $\sim 2.7 (M/M_\odot)\ {\rm AU}$ \citep{kennedy_snowline},
as indicated in Figure~\ref{fig-m_v_snow}, then
the planet orbits at about twice the distance of the snow line, similar to 
the position of Jupiter in our own solar system.  However, the planet's mass of
$\sim 10\mearth$ is close to the critical mass predicted by core
accretion theory \citep{thommes08} where it has exhausted the nearby 
supply of solid material and begins the slow, quasistatic gas accretion phase.
So, MOA-2009-BLG-266Lb fits the theoretical predictions for a large
population of ``failed gas giant" core \citep{laughlin04} planets, which
would have had their growth terminated by the loss of gas from the
proto-planetary disk prior to the rapid gas accretion phase.
Indeed, the distribution of planets found by
microlensing, shown in Figure~\ref{fig-m_v_snow}, seems to confirm this prediction,
as the detection efficiency corrected planetary mass function rises 
steeply, as $\sim q^{-0.7\pm 0.2}$ toward lower mass ratios \citep{sumi10}.
However, a much sharper comparison to theory can be made with
mass measurements of these planets and their host stars. Some theoretical
treatments suggest a relatively sharp feature in the mass function at
$\sim 10\mearth$, and the low-mass end of the exoplanet mass 
function is likely to depend on the host star mass. 
MOA-2007-BLG-192Lb \citep{moa192_naco} is the only other cold, low-mass
planet with a host star mass measurement, but the planet mass is weakly
constrained, due to a poorly sampled light curve.

\section{Conclusions and Implications for Future Discoveries}
\label{sec-conclude}

\begin{figure}
%\centering
%\includegraphics[width=15cm]{mple15_vs_snow11Kss266}
\plotone{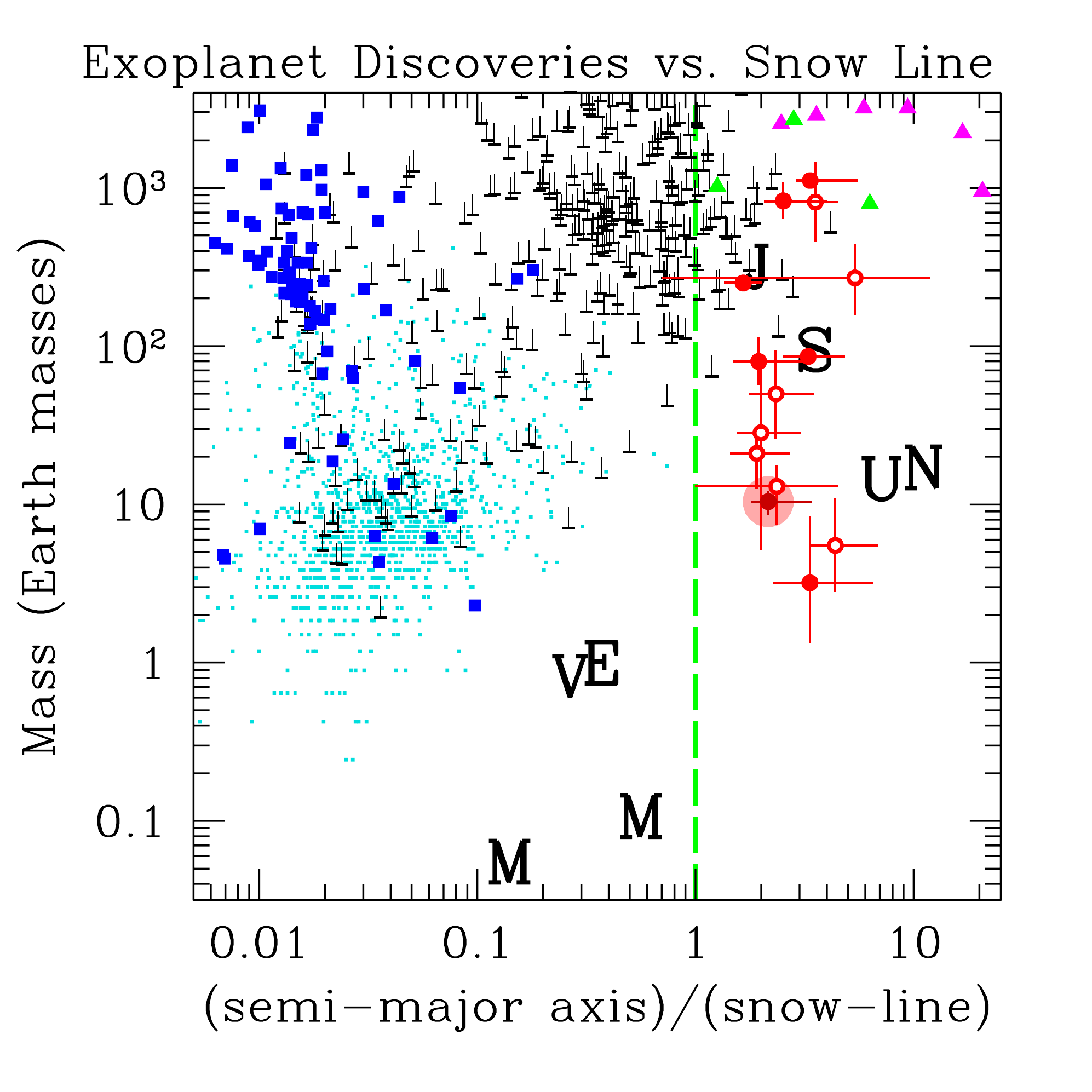}
\caption{The masses of the known exoplanets are shown as 
a function of their semi-major axis divided by the snow line, which is 
assumed to depend on the host star mass as $\sim 2.7M/\msun\,$AU.
Red error-bar crosses indicate microlensing discoveries, with 
MOA-2009-BLG-266Lb indicated by the dark red spot surrounded by the
light red halo. The black, 
upside down ÒTÓs and blue squares indicate exoplanets discovered by 
the radial velocity and transit methods, and the magenta and green 
triangles are planets discovered by imaging and timing. The small 
cyan-colored dots are planet candidates found by Kepler, but not 
yet validated or confirmed (using the mass radius relation of
\citet{traub11}). 
The microlensing planets indicated by open circles have had their 
masses and semi-major axes estimated by a Bayesian statistical analysis.
}
\label{fig-m_v_snow}
\end{figure}

Figure~\ref{fig-m_v_snow} shows the distribution of known exoplanets as a function of mass
and separation, with the separation given in units of the snow line, which is
estimated to be located at $\sim 2.7M/\msun\,$AU (Ida \& Lin 2004; Kennedy,
private communication). (We correct the Ida \& Lin formula to use scale with
the stellar luminosity at the time of planet formation, $\sim 10^6\,$yrs, instead
of the main sequence luminosity.)
The small cyan-colored dots in this plot indicate the location of the
$\sim 1200$ planet candidates recently announced by the Kepler
mission \citep{borucki11}. However, these planet candidates have only radius 
estimates and no mass estimates, so we estimate their masses using
the mass-radius formula of \citet{traub11}. 

While there are a number of exoplanets found by microlensing with similar
mass and separation, MOA-2009-BLG-266Lb is the only low-mass planet
from microlensing with a precisely measured mass.  MOA-2007-BLG-192Lb
is likely to be the lowest mass planet, at $\sim 3\mearth$, found by 
microlensing \citep{bennett08,moa192_naco}, and the mass of the
host star has been reasonably well determined,
$0.084{+0.015\atop -0.012}\msun$ due to a microlensing
parallax signal and detection of the host star in high resolution
AO images. However, the event was not alerted until the planetary
signal was over, and as a result, the planetary light curve is poorly
sampled. This results in an uncertain planetary mass ratio, so that the
mass is not precisely measured. 

Current and future developments in the microlensing field suggest
that such mass measurements may become much more common
in the near future. The rate of microlensing planet discoveries is
expected to increase significantly in the near future, with the 
high cadence, wide-field approach of MOA-II being adopted by
a number of other observing programs, such as the OGLE-IV 
and Wise Observatory surveys, which should begin
full operations in 2011 and 2012, respectively.
The most ambitious ground-based program, the
Korean Microlensing Telescope Network (KMTNet) is
expected to follow a few years later \citep{kmtnet}. 
The observations from the EPOXI
spacecraft have made only a modest contribution to this discovery,
due to the limited observing time and relatively small distance ($\sim 0.1\,$AU)
from Earth. But future observations from EPOXI
or other solar system exploration spacecraft at a more typical
separation of $\simgt 1\,$AU would be much more effective, and will be
able to determine masses for most of the events that they observe.
Finally, follow-up images with the Hubble Space Telescope will
enable mass determinations of many of the planets discovered by microlensing,
after a few years of lens-source relative proper motion \citep{bennett07}.
The results presented here illustrate that it will often be
possible to precisely determine the host star and planet masses, and so
measure the mass function of cold planets in the Earth-Jupiter mass
range as a function of their host mass, which together with the
Doppler and transit methods will provide crucial constraints on the
physics of planet formation across the wide range of planet/star
separations.

This discovery tends to confirm the earlier claims \citep{sumi10,gould10} 
that microlensing has revealed a previously undetected population
of cold, relatively low-mass planets, and the measurement of the
planet and host star masses suggests that this population of planets
may be related to the ``failed Jupiter-cores" predicted by the 
core accretion theory, although there are alternative
mechanisms that could form such planets \citep{boss06}.
Microlensing observations could provide much sharper tests of these
theories if there were more discoveries with precisely measured masses.

One potentially promising avenue for such measurements is 
further observations with small telescopes on 
solar system exploration spacecraft, such as we have obtained 
with EPOXI. While the EPOXI observations of MOA-2009-BLG-266
have made only a modest contribution to the microlensing
parallax measurement, this is a consequence of the poor light
curve coverage of the EPOXI observations and the close proximity
of EPOXI to Earth ($\sim 0.1\,$AU) at the time of observations. Observations
from EPOXI or a similar spacecraft at a more typical ($\simgt 1\,$AU) 
distance, with better light curve coverage (as might be obtained during
an extended mission) would be very effective at measuring lens masses.
Furthermore, such a spacecraft could measure masses for planets and
their host stars residing in the Galactic bulge, which is
probably the case for OGLE-2005-BLG-390Lb
\citep{ogle390} and MOA-2008-BLG-310Lb \citep{janczak10}. These events 
have such short timescales that the orbital motion of the Earth is very unlikely
to allow the measurement of the microlensing parallax, but in most cases,
a telescope in a heliocentric orbit at $\simgt 1\,$AU will be able to measure
the microlensing parallax effect and determine the planet and host star masses.

Of course, the study of low-mass planets beyond the snow line would
benefit greatly from an increased discovery rate over the current rate of
$\sim 4$ per year. The original strategy for finding planets by microlensing
\citep{mao91,gouldloeb92} was to have one wide field-of-view telescope
identify microlensing events that are then observed by a global network of 
narrow field-of-view follow-up telescopes. This strategy was developed in
1992 and was expected to find 
Jupiter-mass planets in Jupiter-like orbits. It has proved not to be
very efficient for lower mass planets, although some important discoveries
have been made \citep{ogle390}.

The development of the high-magnification strategy \citep{griest98,rhie00}
led to a significant increase in observing efficiency because, while
high magnification planetary signals are rare compared to low magnification
planetary signals, the detection efficiency during the brief period of high
magnification is extremely high. Also, small telescopes can do precise
photometry at high magnification. Thus, a substantial fraction of the planet
microlensing planet discoveries to date have come from high magnification
events \citep{gould10}.

However, even when this high magnification strategy is adopted, the
vast majority of the $\sim 700$ microlensing events observed per year
are not effectively monitored for planetary signals. It is simply impossible to
get precise photometry on so many events using narrow field-of-view
telescopes. In order to improve the planetary discovery event rate, it
is necessary to monitor the entire bulge with a high enough cadence so
that planetary signals can be detected in all of the observed microlensing
events, even those without a high planet detection efficiency. There are so
many of these low efficiency events, that they will dominate the total planet
detection efficiency when they can all be monitored.

The MOA-II survey is the first microlensing survey with a large enough
field of view for a high cadence survey. The $\sim 2.2\,{\rm deg}^2$
MOA-II telescope field of view is 
able to observe $13\,{\rm deg}^2$ of the central Galactic bulge
every 15 min, another $13\,{\rm deg}^2$ of somewhat lower priority
bulge fields every 47 min, and $18\,{\rm deg}^2$ of outer bulge
fields every 95 minutes. This has enabled the survey detection
of five planets to date: MOA-2007-BLG-192Lb
\citep{bennett08}, OGLE-2007-BLG-368Lb
\citep{sumi10}, MOA-2009-BLG-266Lb presented here, plus
two additional events from the 2010 season that are under analysis. 
These events (except for MOA-2007-BLG-192) required
detection in real time in order to obtain good coverage of the planetary
anomaly.

We expect the rate of these survey discoveries to increase
quite rapidly in the near future as the number of telescopes involved
in these high cadence surveys is increasing quite rapidly. The
OGLE-IV survey with a $1.4\,{\rm deg}^2$ camera has just begun
on the OGLE 1.3m telescope in Las Campanas, Chile. Although OGLE-IV
has a smaller telescope and field of view than MOA-II, it has better seeing,
and so it should have higher planet detection sensitivity. Nearly complete
longitude coverage should also be possible for part of the season as
a group from Tel-Aviv University is beginning a dedicated Galactic bulge
monitoring program with a $1.0\,{\rm deg}^2$ imager on the 1.0m telescope
at Wise Observatory in Israel in 2011 after a 6-week pilot program in 2010.

The most ambitious project is the Korea Microlensing Telescope Network
(KMTNet) \citep{kmtnet}, which is building a network of
three 1.6 m telescopes equipped with $4.0\,{\rm deg}^2$ cameras
in South Africa, (northern) Chile, and Australia. The KMTNet system will have
the capability for continuous coverage of all bulge microlensing
events by itself, when the weather permits, but it is also locating its
telescopes at different sites from the existing MOA-II, OGLE-IV, and Wise 
telescopes, so that complete light curve coverage will often be possible
when some sites have bad weather.
This should result in a significant increase in the
rate of microlensing planet discoveries.
%\bibliography{scibib}

\acknowledgments
Acknowledgements: We acknowledge the following support:
NASA-NNX10AI81G, NSF AST-0708890 and AST-1009621 (DPB);
National Research Foundation of Korea 2009-0081561 (CH); 
JSPS18253002 and JSPS20340052 (F.A.);
Czech Science Foundation grant GACR P209/10/1318 (DH);
NSF Graduate Research Fellowship (JCY);
European Research
Council Advanced Grant No. 246678 (A.U.);
ESO Prog.ID 385.C-0797(A)

%\begin{figure}
%\centering
%%\includegraphics[width=15cm]{mcmc_rtE_png}
%\includegraphics[width=10cm]{mcmc_pi_E6_png}
%\caption{\noindent 
% Distribution of $\rtEbold$ (or $({\tilde r}_{{\rm E},N},{\tilde r}_{{\rm E},E})$)
%Distribution of the microlensing parallax vector 
%$\piEbold = (\pi_{{\rm E},N},\pi_{{\rm E},E})$ 
%values that are consistent with the
%observed light curve taken from our MCMC runs in the regions of
%the two local $\chi^2$ minima with $u_0 < 0$ and $u_0 > 0$. The
%points are color coded based on their difference from the
%$\chi^2$ minimum of 6053.5, with points with $\Delta\chi^2 < 1$, 4,
%9, 16, 25, and 36 represented by black, red, green, cyan, and gold
% colored dots, respectively. The gray dashed circles indicate
% (approximate) contours of constant $\pi_E = 1/\rep$ and therefore
% constant mass. These assume the best fit  angular Einstein radius,
% $\theta_E = 1.00\,$mas. The black, red, and green points are
% all from the vicinity of the best fit $u_0 < 0$ model, while the 
%cyan and gold points are primarily from the $u_0 > 0$
%solution.
%}
%\label{fig:lightcurve}
%\end{figure} 

\clearpage
\begin{deluxetable}{lrrrr}
\tablecaption{Fit $\chi^2$ values\label{tab-chi2}}
\tablewidth{0pt}
\tablehead{
\colhead{Passband}  & \colhead{data points} & \colhead{best orb.\ fit} & \colhead{best no orb.\ fit} & \colhead{lin.-LD no orb.\ fit}  }
\startdata
MOA-red$^*$ & 1996 & 1983.79 & 1981.82 & 1983.57  \\
FTS-SDSS-$I^*$ & 128 & 121.15 & 127.36 & 130.52  \\
Canopus-$I^*$ & 205 & 205.84 & 204.46 & 205.96  \\
Wise-$I^*$ & 36 & 34.33 & 34.86 & 33.40  \\
Wise-$R^*$ & 30 & 27.82 & 27.83 & 27.84  \\
Bronberg-un$^*$ & 597 & 601.92 & 601.01 & 601.57  \\
IRSF-$K^*$ & 18 & 16.68 & 16.68 & 16.54  \\
IRSF-$H^*$ & 20 & 20.13 & 21.00 & 19.50  \\
IRSF-$J^*$ & 19 & 19.15 & 19.79 & 19.35  \\
SAAO-$I$ & 33 & 32.58 & 32.59 & 32.60  \\
CTIO-$H^*$ & 861 & 862.57 & 862.26 & 853.64  \\
CTIO-$I^*$ & 317 & 309.42 & 308.68 & 309.61  \\
CTIO-$V^*$ & 56 & 56.08 & 56.08 & 56.49  \\
Danish-$I^*$ & 611 & 603.66 & 603.25 & 610.11  \\
OGLE-$I$ & 929 & 920.60 & 920.70 & 920.70  \\
Mt.~Lemmon-$I$ & 73 & 71.01 & 70.97 & 71.26  \\
FTN-SDSS-$I^*$ & 148 & 144.76 & 145.18 & 149.17  \\
EPOXI-un & 21 & 18.68 & 18.88 & 18.87  \\
total & 6098 & 6050.17 & 6053.41 & 6060.68  \\
\enddata
\tablecomments{ 
$^*$ denotes passbands using limb darkening tables. ``un" denotes unfiltered. }
\end{deluxetable}

\begin{deluxetable}{lccccc}
\tablecaption{Best Fit Parameters\label{tab-fitpars}}
\tablewidth{0pt}
\tablehead{
\colhead{Parameter}  & \colhead{units} & \colhead{best-orb.} & \colhead{MCMC orb.} & \colhead{best no orb.} & \colhead{MCMC no orb.} }
\startdata 
$\chi^2$/dof & & 6050.17 & & 6053.41 & \\
$t_E$ & days & 61.447 & $61.47\pm 0.40$ & 61.612 & $61.54\pm 0.36$ \\
$t_0$ & HJD$'$ & 5093.257 & $5093.257\pm 0.083$ & 5093.298 & $5093.285\pm 0.051$  \\
$u_0$ & & 0.13158 & $0.1315\pm 0.0008$ & 0.13129 & $0.1314\pm 0.0008$ \\
$s$ & & 0.91429 & $0.91434\pm 0.00036$ & 0.91425 & $0.91421\pm 0.00036$ \\
$q$ & $10^{-5}$ & $5.815$ & $5.63\pm 0.25$ & $5.425$ & $5.45\pm 0.07$\\
$\theta$ & rad & 2.2677 & $2.2692\pm 0.0034$ & 2.2715 & $2.2708\pm 0.0035$ \\
$t_\star$ & days & 0.33140 & $0.3262\pm 0.0068$ & 0.32152 & $0.3216\pm 0.0012$ \\
$\pi_E$  & & 0.2094 & $0.223\pm 0.037$ & 0.2395 & $0.232\pm 0.038$ \\
$\phi_E$ & rad & 0.760 & $0.74\pm 0.16$ & 0.640 & $0.70\pm 0.14$ \\
$\pi_{E,N}$ & & 0.1517 & $0.1665\pm 0.0503$ & 0.1921 & $0.1795\pm 0.0493$ \\
$\pi_{E,E}$ & & 0.1442 & $0.1439\pm 0.0035$ & 0.1430 & $0.1435\pm 0.0035$ \\
$\dot s_x$ & $10^{-3}\,$day$^{-1}$ & $-0.34$ & $-0.25\pm 0.15$ & & \\
$\dot s_y$ & $10^{-3}\,$day$^{-1}$ & $-2.05$ & $-0.84\pm 1.49$ & & \\
$T_{\rm orb}$ & days & 2380 & & & \\
\enddata
\tablecomments{ Parameters are given in an inertial geocentric frame fixed at 
${\rm HJD}' = 5086$
(${\rm HJD}'={\rm HJD}-2450000$)}
\end{deluxetable}

\begin{deluxetable}{cccc}
\tablecaption{Physical Parameters\label{tab-pparam}}
\tablewidth{0pt}
\tablehead{
\colhead{Parameter}  & \colhead{units} & \colhead{value} & \colhead{2-$\sigma$ range} }
\startdata 
$D_L $ & kpc & $3.04\pm 0.33$ & 2.4-3.7 \\
$M_\star$ & $\msun$ & $0.56\pm 0.09$ & 0.39-0.74 \\
$m_p$ & $\mearth$ & $10.4\pm 1.7$ & 7.2-14 \\
$a$ & AU & $3.2{+1.9\atop -0.5}$ & 2.3-13  \\
$P$ & yr & $7.6{+7.7\atop -1.4}$ & 5.4-62 \\
\enddata
\tablecomments{ Uncertainties are
1-$\sigma$ parameter ranges. }
\end{deluxetable}

\end{document}